\documentclass[preprint,12pt,authoryear]{elsarticle}

\usepackage{amssymb}

\usepackage{xcolor}

\journal{Icarus}

\newcommand{\be}{\begin{equation}}
\newcommand{\ee}{\end{equation}}
\newcommand{\beq}{\begin{eqnarray}}
\newcommand{\eeq}{\end{eqnarray}}

\begin{document}

\begin{frontmatter}



\title{Apse-alignment in narrow-eccentric ringlets and its
implications for the $\epsilon$-ring of Uranus and the ring system of (10199) Chariklo}


\author{Melita.M.D.$^{1,2}$, Papaloizou J.C.B.$^3$}

\address[$^{1}$]{Instituto de Astronom\'ia y F\'isica del Espacio.
(CONICET-UBA). Intendente G\"uiraldes S/N. CABA. C1428ZAA. Argentina.}
\address[$^{2}$]{Facultad de Ciencias Astron\'omicas y Geof\'isicas.
Universidad Nacional de La Plata. Paseo del Bosque S/N. La Plata.
B1900FWA. Argentina.} 
\address[$^3$]{DAMTP, University of Cambridge, CMS, Wilberforce Road,
Cambridge CB3 0WA, UK.}

\begin{abstract}

The discovery of ring systems around objects of the outer Solar System
provides a strong motivation to apply theoretical models in order to
better estimate their physical and orbital parameters, which can 
constrain scenarios for their origin.

We review the criterion for maintaining apse-alignment  across a ring
and the balance between the energy input rate provided by a close by 
satellite and the internal dissipation rate occurring through ring
particle collisions that is required to maintain ring eccentricity, as
derived from the equations of motion governing the
Lagrangian-displacements of the ring-particle orbits. We use the case
of the $\epsilon$-ring of Uranus, to calibrate our theoretical
discussion and illustrate the basic dynamics governing these types of
ring. 

In the case of the ring system of (10199) Chariklo, where the evidence
that the rings are eccentric is not conclusive, we apply the theory of
apse-alignment  to derive information about the most plausible
combination of values of the surface density and
eccentricity-gradient, as well as the masses and locations of their
postulated but -presently undetected- shepherd-satellites.  

When the balance conditions that we predict are applied to the ring
system of (10199) Chariklo, we are able to estimate the minimum mass
of a shepherd satellite required to prevent eccentricity decay, as a
function of its orbital location, for two different models of
dissipation. We conclude that the satellite mass required to maintain
the $m=1$ eccentric mode in the ring, would be similar or smaller than
that needed to confine the rings radially.

Our estimation of the most plausible combinations of 
eccentricity gradient and surface density  consistent with apse-alignment  are based on a standard
model for the radial form of the  surface  density distribution, which  approximately
agrees with the optical depth profile derived by the stellar
occultations. %
 We find a diverse range  of solutions, with combinations of
eccentricity gradient and surface mass density that tend to minimize required 
enhanced collisional effects,  having adopted estimated values of
the form factor of the
second degree harmonic of the gravitational potential.

\end{abstract}


\end{frontmatter}


\section{Introduction}
\label{sec.Intro}

The discovery of the ring system around the Centaur (10199) Chariklo
\citep[][hereafter BR14]{Braga-Rivasetal.2014} , the trans-Neptunian
object (136108) Haumea  \citep{Ortizetal.2017} and the evidence that (2060) Chiron
\citep{Ortizetal.2015} is also surrounded by rings, poses an
interesting theoretical challenge, not only to explain the origin of
these unexpected features, but also to estimate better their physical
and orbital parameters. 

The surface density and the orbital frequency of the ring particles
make the Centaur's rings remarkable analogs of the narrow rings of
Uranus or the outer rings of Saturn  (BR14),
 However, it is not apparent that the same  formation  mechanism  has occurred 
in all cases.

A number of scenarios have been put forward to explain the origin the
rings, i.e., cometary-like activity \citep{PanWu2016}, disruption of
an asteroidal body through a close encounter with a major planet
\citep{Hyodoetal.2016}, ejecta produced in a physical collision either
on the surface of the Centaur or with a conveniently located satellite
\citep{Melitaetal.2016} and the  disaggregation of a satellite caused by
the Centaur's tidal field \citep{Melitaetal.2016}. In any case, it is
apparent that a more precise estimation of basic physical and/or
orbital parameters than is currently available would be helpful in
deciding between different formation scenarios.

It has been shown that the rings of (10199) Chariklo are stable under
close encounters with the major planets \citep{Araujoetal.2016}. On
the other hand, their migration timescale induced by the 
Poynting-Robertson effect is approximately $10^7 yr$ and the spreading
time scale arising from energy loss through inelastic collisions that
result in an effective viscosity is only about $10^6 yr$ (BR14). These
short time scales indicate that the rings are being confined in their
present location.  A general argument can be made that
such confinement requires the injection of energy with zero net angular momentum
that is most likely supplied by non axisymmetric forcing \citep[eg.][]
{Longaretti2017} which could originate from shepherd satellites.
However, although this is viable in the case of the $\epsilon$ ring of Uranus
no such shepherds have been identified for the $\alpha$ and $\beta$ rings
\citep[but note] [who suggest that they may be below the limits of detection]{Hedman2016}
and many other ringlets around Saturn.  Such satellites could also
account for the maintenance of ring eccentricity. In this situation it is natural that other
phenomena  have been suggested as providing required forcing.
These include normal oscillation modes of the planet in the case of the
Maxwell ringlet \citep[][]{Frenchetal.2016}  and non axisymmetric variations
of the B ring in the case of ringlets in the Cassini division \citep[][]{Hedmanetal2010}.

 Thus the rings of (10199) Chariklo  may be confined
 by shepherd satellites, estimated to have a physical
radius of about $1 km$ with an assumed bulk-density of $1 g\ cm^{-3}$.
 We shall assume that such a satellite  may also act to maintain ring eccentricity, however 
we note that the issue
of the maintenance of uniform apsidal precession  discussed below does not require this and 
would operate if some other phenomena provided an equivalent ring forcing.

Continuous observations of stellar occultations by (10199) Chariklo 
should enable us to better constrain the geometry of the ring system.
At the present time, available observations are still compatible with
a circular shape, with more being needed to establish its eccentric
nature \citep{Berardetal.2017}. Nonetheless, the similarity between
the asteroidal ringlets and the eccentric ringlets of the planets
Saturn and Uranus indicates that the ones surrounding Centaurs may
also be eccentric. In that case the orbits of the ring-particles must
have their peri-apses alined and be precessing uniformly. A fact that
makes it plausible that one of the centaur's rings posesses a
non-neglibile eccentricity is that the spin-orbit 1:3   second order resonance
between the mean orbital frequency in a circular orbit and the
intrinsic rotation frequency of (10199) Chariklo   that  could be  manifested  through a mass anomaly
on it  or within it   rather than 
a 2:6 fourth-order resonance resulting from its ellipsoidal 
shape
is close to or within
the innermost edge of the ring system.  These higher order resonances 
 could play a relevant role in maintaining the eccentricity of the
rings as well as providing energy and angular momentum input to
counteract orbital decay resulting from collisional dissipation in a
similar manner to shepherd satellites \citep[see][]{SICARDYETAL2018} .

The precession of a ring particle orbit is induced by the oblateness
of the primary with the angular velocity of the apsidal line being
given, in the limit of very small eccentricity by
\citep{MurrayandDermott} as
$$ { \dot{\varpi} } (r_0) = \frac{3}{2}\ J_2\ \Omega
\left(\frac{R}{r_0}\right)^2, $$ 
where  $\varpi$ is the longitude of periastron and $\dot{\varpi}$
its time derivative, the physical radius of the primary is $R$, its mass  being $M_0$, the
form-factor of the second degree harmonic of its gravitational
potential is $J_2$, $r_0$ is the mean-distance between the particle
and the primary, $\Omega= \sqrt{{G\ M_0} / r_0^3}$, and $G$ is the
gravitational constant.  Since  $  \dot{ \varpi  }(r_0)$  decreases
with increasing values of $r_0$, in a ring system that is precessing
uniformly, the differential precession that would be induced by the
oblateness of the primary, if it alone acted, must be somehow
compensated.

The leading model that accounts for uniform precession in narrow
eccentric ringlets is the "self-gravity model''
\citep{GoldreichandTremaine1979}, since precession induced by the
self-gravity of the ring opposes that induced by the oblateness of the
primary.  A signature of narrow rings dominated by self gravity is
that the eccentricity gradient, $ a\partial e/\partial a,$ with $a
\equiv r_0$ being the semi-major axis, is a positive quantity such
that the orbits of the inner ring particles are less eccentric than
the outer ones (see \citep[][hereafter PM05]{PapaloizouandMelita2005}.
Several ringlets in the Solar System are known to exhibit this type of
behavior, for example the Titan and Maxwell rings of Saturn
\citep{Porcoetal.1984}, and the $\alpha$, $\beta$ and $\epsilon$ rings
of Uranus \citep{Elliotetal.1977}.   However,
\citet{GoldreichandPorco1987} find that, a model which balances
differential precession induced by the central object with effects due
to self-gravity, produces mass estimates for the $\alpha$, $\beta$
rings that are too low to be consistent with the possibility of
confinement by perturbing satellites as well as high measured $s$ band
radio opacities. These indicate that the masses are much larger,
leading to the possible presence of collisional effects that counter
the effects of increased self-gravity, as we indicate below ( see also
\citet{Longaretti2017} for further discussion).
Collisional interactions such as those associated with close packing
arising from converging streamlines and perturbations due to any
shepherd satellites should   be taken into account
\citep{BorderiesGoldreichandTremaine1983, MosqueiraandEstrada2002,
ChiangGoldreich2000}.  We remark that the importance of the fact that the
particles are close-packed at pericenter for the dynamics was
recognised at an early stage \citep{DermottandMurray1980}.  

Here, we review a theoretical model of narrow-eccentric ringlets 
that has been applied to the rings of Uranus and the Titan-ringlet of
Saturn \citep[PM05 and ][respectively]{MelitaandPapaloizou2005} Our
aim is to investigate how ring eccentricity is maintained against 
collisional dissipation and different processes are able to contribute
to the maintenance of uniform precession. We  consider the
importance of significant collisional interactions resulting from
phenomena such as close packing, intersecting streamlines
  and torquing from external satellites together with 
the associated energy loss. In addition we   relate this to shepherd satellites
postulated to be needed to counteract ring spreading and orbital decay.

When  applied to the case of the
ring system of (10199) Chariklo, this discussion  leads to 
constraints on the masses and orbits of putative shepherd satellites,
the mass density of the rings and  collisional interactions that
occur between  ring particles.

In section \ref{sec.Ch} we review the latest estimates of the
physical and dynamical properties of (10199) Chariklo and  its rings.
In section \ref{sec.Theory} we review the apse-alignment 
theory, deriving the conditions to maintain a non-zero mean
eccentricity and the apse-alignment  of the ring particle
orbits.  

\noindent We apply the radial-action balance equation to estimate the mass of
the shepherd satellites responsible for maintaining the eccentricity
 in section \ref{CH_ring}. 
In section \ref{aps-al}
we validate the equation that expresses the balance of impulses to achieve 
the apse alignment  for the case of the $\epsilon$-ring of Uranus and
apply it to the ringlets of (10199) Chariklo. In the last section we
discuss our results and draw some conclusions.

\section{(10199) Chariklo and its ring system}
\label{sec.Ch}

The main orbital and physical properties of Centaur (10199)
Chariklo are summarized in table \ref{tab.1}. Its orbit is fully
contained between the orbits of  Saturn and Uranus, with
moderate values of eccentricity and inclination. It is known that the
Centaur population exhibits a bimodal distribution of surface colour
indexes \citep{Peixinhoetal.2003} and (10199) Chariklo belongs to the
``grey'' or ``neutral'' group (as opposed to the ``red'' group), which
might be associated with the existence of cometary-like activity in the
past \citep{MelitaLicandro2012}.  
\begin{table}[!h]
\begin{tabular}{|l|c|}
\hline

\hline
Absolute magnitude$^{(1)}$ & $6.6$ \\
\hline
Apparent magnitude$^{(1)}$ (V) & $18.5$ \\
\hline
Colour index$^{(1)}$ ($|B-V|$) & $0.86$ \\
\hline\
Orbital semimajor axis$^{(1)}$ & $15.75$ AU\\
\hline
Orbital eccentricity$^{(1)}$ & $0.15$ \\
\hline
Orbital inclination$^{(1)}$ & $23$ deg\\
\hline
Physical radius (Jacobi)$^{(2)}$ R$_{1J}$  & ($157 \pm 4$) km\\
\hline
Physical radius (Jacobi)$^{(2)}$ R$_{2J}$  & ($139 \pm 4$) km\\
\hline
Physical radius (Jacobi)$^{(2)}$ R$_{3J}$  & ($86 \pm 1$) km\\
\hline
Bulk Density (Jacobi)$^{(2)}$,  $\rho_J$  & $0.796^{+4}_{-2}\ g\ cm^{-3}$\\
\hline
Albedo (Jacobi)$^{(2)}$, $p_V$ & $0.042 \pm 0.1$ \\
\hline
Mass (Jacobi)$^{(2)}$ &  $(6.1\ \pm 0.1)\ 10^{21}$ g\\
\hline
Sidereal intrinsic rotation period$^{(2)}$ & ($7.004 \pm 0.001$) hr \\
\hline
\end{tabular}
\caption{
$^{(1)}$ Physical properties and orbital properties of  (10199)
Chariklo taken from {\it https://ssd.jpl.nasa.gov/}. 
$^{(2)}$ Posterior values of the physical parameters of (10199)
Chariklo modelled as a Jacobi 
ellipsoid with the given intrinsic rotational sidereal period, from
\citet{Leivaetal2017}. }
\label{tab.1}
\end{table}
In Table  \ref{tab.1} we only quote  the estimate of the shape of the body
of (10199) Chariklo made assuming it to be  a Jacobi ellipsoid,  noting that  those made 
assuming  an arbitrary triaxial ellipsoid are similar. 
We note that, according to the most recent  reference
\citep{Leivaetal2017}, the oblateness
 is remarkably large, $\epsilon \approx 0.45 \pm 0.1$.
 Based on  the oblateness of the body, the  form factor of the second
degree harmonic of the gravitational potential, $J_2$ can be
approximated as 
$$J_2^{(o)} = \frac{1}{5}\ \epsilon\ (2-\epsilon)$$
\citep{PanWu2016}. On the other hand if it is assumed that the shape
of the body is given by the equipotential surface   determined by the
self-gravity and the centrifugal potential, the form factor of the
second degree harmonic of the gravitational potential is given by
$$J_2^{(e)} = \frac{2}{3}\ \epsilon - \frac{q}{3},  \hspace{3mm}  {\rm where}$$ 
$$q =  (f^2 R^3)/(G M_0),$$ 
with $f$ being the intrinsic rotational frequency, $G$ the gravitational
constant, $R$ is the physical mean radius of the body and $M_0$ its mass
\citep{MurrayandDermott}. We find that if it is assumed that (10199)
Chariklo is a Jacobi ellipsoid, then $J_2^{(e)} = 0.13$ and $J_2^{(o)} =
0.28$, whereas if an arbitrary ellipsoidal shape is assumed,
$J_2^{(e)} = 0.11$ and $J_2^{(o)} = 0.18$. It must be noted that
these estimations of $J_2$ are about two orders of magnitude larger than the
typical value for a major planet in the Solar System and also that the
early estimate, made assuming that the values of the radii of the body
are those given in BR14,
render a smaller value of $J_2 \approx 0.076$  \citep[see
also][]{PanWu2016}.

The discovery of a ring system consisting of two dense ringlets separated
by a gap was made through the observation of multi-chord stellar
occultations (BR14). The two ringlets are known as ``CR1'' and
``CR2''. A recent 
update on the physical properties of the  (10199) Chariklo ring system has been
reported by \citet{Berardetal.2017} (B17) where it is 
argued that some of the properties of the CR1 ringlet (shaped profile,
variation of width with longitude and sharp edges) make it 
quite  similar to those of the narrow eccentric ringlets of Saturn 
\citep{  Nicholsonetal.2014, Frenchetal.2016} or Uranus
\citep{ElliotandNicholson1984, Frenchetal.1986} .
 
In Table \ref{tab.2} we summarize some of
the ring properties reported in B17.  It has been estimated that 
the total eccentricity variation across the ring, may be of the order of 
$\delta e = {\delta W}/({2r_0})  \sim 0.003,$ 
 with ${\delta W}$ denoting the width variation, 
which we also asssume equal to the lower limit of the    
eccentricity  at the inner edge.
 However, the most statistically
significant model of the ring, performed using
the latest data available at present is compatible  with a circular ring. 

In summary, for our current purposes  we  shall assume an   inner  eccentricity,
$e=0.003$,  with
a constant eccentricity gradient that is compatible with an eccentricity  variation of
$\delta e =0.003$, as estimated in B17.  To obtain a more accurate
estimation of the geometry in the future, the degeneracy between the
ring eccentricity and the position of its pole has to be lifted,
which should  occur with the accumulation of future observations of
stellar occultation events. 
\begin{table}[!h]
\begin{tabular}{|l|c|}%
\hline
CR1 ringlet radius, r$_{CR1}$ & 390.6 $\pm$ $3.3$ $km$ \\
\hline
CR2 ringlet radius, r$_{CR2}$ & 405.4 $\pm$ $3.3$ $km$ \\
\hline
Optical Depth of the CR1  ringlet,  $\tau_{CR1}$ & $0.45$ \\
\hline
Optical Depth of the CR2 ringlet, $\tau_{CR2}$ & $0.05$ \\
\hline
Width of the CR1 ringlet, $W_{CR1}$ & ($5.5-7.0$) $km$ \\
\hline
Width of the CR2 ringlet, $W_{CR2}$ & ($0.1-1.75$) $km$ \\
\hline
Width variation of the CR1 ringlet,  $\delta W$ & $\pm 3.3 km$ \\
\hline
Width of the gap,  $W_{Gap}$ & $\approx 14.8$ $km$ \\
\hline
Sharpness of the CR1 ringlet, $S_{CR1}$ & $<<1$ $km$ \\ 
\hline
Optical Depth in the gap,  $\tau_{Gap}$ &   $< 0.004$ \\
\hline
%
%

Mean orbital Period of the rings,  $<T>$ & $\sim 0.79\ d $ \\
\hline
Mean orbital frequency, $<\Omega>$ &  $\sim 7.96\ d^{-1} $\\  
\hline

Mean precession frequency, $ <\dot{\varpi} (J_2^{a}=0.28)>$ & $0.34\ d^{-1}$ \\
\hline
Mean precession frequency, $ <\dot{\varpi} (J_2^{b}=0.076)>$ & $0.09\ d^{-1}$ \\

\hline
\end{tabular}
\caption{Physical and orbital parameters of the ring system of 
(10199) Chariklo taken from (B17) and (BR14).}
\label{tab.2}
\end{table}
Relevant evolutionary  timescales for  the system are estimated
to be very short with respect to the dynamical age of a typical Centaur. 
The timescale in which the rings would spread a distance equivalent to
its orbital location by the Poynting-Robertson effect is
estimated  by BR14  to be , T$_{PR} \sim 10^9R$ $yr$ and
through viscous dissipation, assuming the ring to be a monolayer,  to be,  T$_{\nu} \sim 10^4/R^2$ $yr,$
where $R$ is the particle size in metres.
The maximum survival time against orbital dislocation occurs
for $R \sim 0.02$ and is $\sim 2\times 10^7$ $yr.$
This time will be significantly less for significantly  larger or smaller particles. 
 Hence the existence  of shepherd satellites that
provide confinement is implied,  as otherwise the system would have  a 
remarkably  unlikely short age. 
The physical radius of a shepherd satellite capable of
 producing  an adequate   shepherding effect, was estimated to be 
about  $R_{shepherd} \sim 3 km$ for a bulk density of $\rho = 1
g/cm^{3}$. While the physical radius of a body capable of  clearing  the
observed gap between the CR1 and CR2 ringlets is $R_{Sat-Gap}
\sim 1 km$,  for a  bulk-density of $\rho = 1 g/cm^{3}$ (see BR14). 

\section{The dynamics of apse-alignment  and eccentricity maintenance}
\label{sec.Theory}

The goal of the theoretical modelling of the ringlets is to explain
quantitatively how the eccentricity is sustained against collisional
dissipation and how the different forces acting are balanced to
produce rigid precession.  In PM05 the rings were modelled as a two
dimensional continuum adopting a Lagrangian description.  The
equations of motion are developed for perturbations with respect to a
background state in which ring particles are in circular orbits and
are characterised by the associated Lagrangian displacement.

In the axisymmetric unperturbed state, 
particles are in circular motion, such that
$$r = r_0 ,$$ $$ \theta = \theta_0= \Omega(r_0)t + \beta_0.$$
 Here we adopt a cylindrical coordinate system $(r , \theta)$ with
origin at the central mass, $M_0$. Thus the value of, $r,$ for the circular
orbit of a given particle in the background state is $r_0,$
$\Omega(r_0)$ is its angular velocity and $\beta_0$ is a phase factor.
From now on a subscript, $0,$ denotes application to the background
state. The components of the Lagrangian displacement from the
axisymmetric background state are given by ${\bf \xi} = (\xi_r,
\xi_{\theta}).$  These are such that the perturbed coordinates of
a perturbed particle are given by 
\be r = r_0 + \xi_r,\ee and \be r_0
\times (\theta -\theta_0 ) = \xi_{\theta}.\ee

The Lagrangian variation of
given quantity $Q$, $\Delta(Q)$ is defined as
\be
\Delta(Q) = Q\left( {\bf r}_0 +{\mbox{ {\boldmath$ \xi$}} }\right) - Q_0\left( {\bf
r}_0 \right),
\ee
where $Q $ and $Q_0$ are the values of the given physical quantity in the
perturbed and unperturbed flow respectively.
It thus represents the variation as experienced by a fluid element at original
location ${\bf r}_0.$
In contrast, the Eulerian variation, measured at a fixed location,  is defined as
\be
\delta(Q) = Q\left( {\bf r}_0 \right) - Q_0\left( {\bf r}_0 \right).
\ee
Thus  to first order in ${\bf \xi}$ the operators, $\Delta$ and $\delta$ are  related by
\be
\Delta  = \delta +\mbox{ {\boldmath$\xi$}} \  {\bf .} \nabla
\ee
and we note that $\Delta r = \xi_r$ and $\Delta \theta = \xi_{\theta}/r_0.$

\noindent The components of the equation of motion we use  are given by (see PM05)
\be
\Delta \left( \frac{d^2 r}{dt^2}  - r\left(\frac{d\theta}{dt}\right)^2 +\frac{\partial \psi_{M_0}}{\partial r}
  \right)  = {d^2 \xi_r\over dt^2} -2\Omega {d \xi_{\theta} \over dt}+ 2\xi_r r_0
\Omega {d \Omega \over dr_0} = f_{r} - \Delta \left({\partial \psi ' \over
\partial r} \right)
\label{pr}
\ee
\be \Delta  \left( r^2\frac{d^2\theta}{dt^2} +2r\frac{d r}{dt}\frac{d\theta}{dt} \right) = 
{d^2 \xi_\theta \over dt^2} + 2\Omega {d \xi_r \over dt} =
f_{\theta} - \Delta \left( {1\over r} {\partial \psi '\over
\partial \theta } \right), \label{pth0}
\ee
where $\psi_{M_0}$ is potential due to the central mass and 
where $\psi'$ is the sum of the potentials related to the satellites,
$\psi_{s}$ and the
self-gravity of the ring, $\psi_{SG}$, 
$f_r = \Delta(F_{r}), f_\theta = \Delta(F_{\theta})$ are the
variations of the components of the force per unit mass due to particle
interactions.  These could be associated with dissipative collisions  and/or as indicated by \citet{ChiangGoldreich2000}
an effective pressure. 
 In writing  equations~(\ref{pr}) and (\ref{pth0}) we neglect 
  background self-gravity  on account of the small mass of the ring.

 We comment that
in this section we  adopt the time averaged figure of the
central  object. Thus we do
 not consider harmonically  varying gravitational forces
resulting from the  non axisymmetric shape.
Such effects can act in a similar manner and be treated with the same formalism
applicable the perturbations from a shepherd satellite discussed in Section \ref{RST}

  We  further remark that $d/ dt $ denotes the Lagrangian time derivative on a fixed particle.
In addition the perturbing satellites are not included in the background state but are introduced as
perturbing quantities which act on the particles at their perturbed position. Thus
\be \Delta \nabla \psi_s \equiv    \nabla \psi_s |_{{\bf r_0},\theta_0}  +{\mbox{\boldmath$\xi$}}\cdot\nabla (\nabla\psi_s)  |_{{\bf r_0},\theta_0}  . \label{sss}\ee
The second term on the right hand side of (\ref{sss}) is accordingly of second or higher order in perturbed quantities
and would be omitted in a strictly linear analysis.
On the other hand in working with the left hand sides of  equations~(\ref{pr}) and (\ref{pth0})  we make the epicyclic approximation
and accordingly retain only  terms of first order in the perturbations. 
In addition we have $\Omega = r_0^{-1/2} \sqrt{\partial \psi_{M_0}/\partial r_0}.$

\subsection{The condition for rigid-precession }\label{Rigidprec}

We here consider the situation when the perturbation takes
form of  an  $m=1$ mode  which forms a steady pattern in an appropriate rotating frame.
This corresponds to the ring  being  eccentric and precessing uniformly.
The frame in which the pattern is stationary rotates with the precession frequency $\Omega_P.$

To obtain a condition for this to hold,
 we rewrite 
equation~(\ref{pr})  as

\be
{d^2 \xi_r\over dt^2} + \xi_r \kappa^2
= f_{r} - \Delta \left({\partial \psi '\over \partial r} \right)  + 2 \Omega {\cal Q}_\theta
\label{pr2}\ee 
where the quantity ${\cal Q}_{\theta}$ is defined by
\be
{\cal{ Q_{\theta}} } = {d\xi_{\theta}\over dt} +  2\ \Omega \xi_r
\ee
and the square of the epicyclic frequency is given by 
\be
 \kappa^2=4\Omega^2 +2r_0\Omega {d \Omega \over dr_0} .
\ee
We note that the local  precession frequency of a free particle orbit
is  $\dot{\varpi}(r_0)  = \Omega(r_0) -\kappa(r_0).$

Following \citep{S85}, we transform to the coordinate system rotating
with the pattern frequency $\Omega_P.$ In this frame  the
azimuthal angle in the background state becomes $\phi_0 = \theta_0 -
\Omega_P\ t.$ We look for steady solutions in this frame which depend
only on $r_0$ and $\phi_0.$ In this case in equations~(\ref{pr}) -
(\ref{pr2}) we have
\be
\frac{d}{dt} \rightarrow  (\Omega -\Omega_P) \frac {\partial}{\partial \phi_0}.
\ee
Making use of this, after straightforward algebra, we obtain the following relation  from 
these equations 
\newpage
$$
 \frac{1}{2\pi} \int_0^{2\pi}\left( \frac{\kappa^2}{(\Omega - \Omega_P)^2} - 1 \right) \xi_r\cos(\phi_0) d\phi_0 =
  $$
  \be
\frac{1}{(\Omega - \Omega_P)^2 }
\left( F_{cr} + g_D(r_0) + \frac{2 \Omega  F_{c\theta}}{(\Omega - \Omega_P)} \right)
\label{pr3}
\ee

where:
\be
F_{cr} =  \frac{1}{2\pi} \int_0^{2\pi} f_r \cos(\phi_0) d\phi_0,
\ee
is the integrated  radial component contribution of the collisional interaction, 
\be
 F_{c\theta} =  \frac{-1}{2\pi} \int_0^{2\pi} f_\theta \sin(\phi_0) d\phi_0.
\ee
 is the   integrated  azimuthal component  contribution of the collisional
interaction, and 
\be
g_{D}(r_0) = -  \frac{1}{2\pi} \int_0^{2\pi}\left(\cos(\phi_0) \Delta  \left({{\partial
\psi_{SG}} \over {\partial r}}\right) -\frac{2 \Omega \sin(\phi_0)}{(\Omega - \Omega_P)}
\Delta\left(\frac{1}{r}{{\partial \psi_{SG}} \over {\partial \theta}}\right)\right) 
 d\phi_0,
\ee
gives the contribution form the self-gravity of the ring, where we are
assuming that there are no satellites that contribute directly to the
form of the $m=1$ wave mode.


Assuming that the eccentricity is small, so that we may adopt the
epicyclic approximation,
for which the  radial displacement takes the harmonic form  $\xi_r = A(r_0)\ cos(\phi_0),$ 
on integrating over $\phi_0,$ equation~(\ref{pr3}) gives
\be
 \frac{1}{2} \left( \frac{\kappa^2}{(\Omega - \Omega_P)^2} - 1 \right) A(r_0) =
\frac{1}{(\Omega - \Omega_P)^2 }
\left( F_{cr} + g_D(r_0) + \frac{2 \Omega  F_{c\theta}}{(\Omega - \Omega_P)} \right)
\label{pr3a}
\ee
  By using the approximation $\Omega >> \Omega_P$ and $\Omega >>
 \dot{\varpi} $, equation~(\ref{pr3a}) can be written 
\be 
(\Omega_P -  \dot{\varpi} )\ A(r_0) = \frac{g_{ext}}{\Omega(r_0)} 
\label{pr5} 
\ee 
where 
$$ g_{ext} = F_{cr} + g_D(r_0) + 2 F_{c\theta}. $$
Equation (\ref{pr5})  has to be satisfied to enable  rigid precession of the ring.
When external satellites are absent, the latter  occurs when
contributions from  self gravity   and  collisions
on the right-hand side 
balance the differential precession term on the left-hand side.

\subsection{The self-gravity term}

For the computation of $g_D$ we also follow S85. The local self-gravity at
$r_0$ is canonically assumed to be that of a band of thickness $\Delta r =
r_2 - r_1$, where $r_1$ and $r_2$ are the inner and outer limits of the
unperturbed ring respectively.
 In addition we assume the radial scales in the problem are much smaller than the
angular scales so that the $\theta$ component of the force due to self-gravity may be neglected.
This is the tight winding approximation. 
  Thus we write (PM05)
\be
\Delta  \left({\partial \psi_{SG} \over \partial r} \right)  
=\frac{ 2G}{{\bar r}}\ \int_{r_1}^{r_2} \frac{\Sigma(r')r'}{(r -
r')} dr'
\label{gs1p}
\ee
where $G$ is the gravitational constant,  and ${\bar r}$ is the mean radius of the background ring. 
We set   $r = r_0 + \xi_r$ and $r' = r_0' +
\xi_r'$, where we recall $\xi_r = \xi_r(r_0,\phi_0) = A(r_0)\ cos(\phi_0) $ and $\xi_r' =
\xi_r(r_0',\phi_0) = A(r_0')\ cos(\phi_0)$. 
 We remark that 
 the background
potential due to self-gravity is included in (\ref{gs1p}). However, this is inconsequential 
as  it has been considered negligible.   
In addition  principal values of the  integrals are to be taken.

 In  the tight-winding approximation the conservation of mass  of a moving fluid element 
requires that
\be
r'\Sigma(r')dr' = \Sigma(r_0')r_0'  dr_0' 
\ee
 Thus we have
\be
\Delta  \left({\partial \psi_{SG} \over \partial r} \right)
= \frac{2G}{{\bar r}}\ \int_{r_1}^{r_2} \frac{\Sigma(r_0')r_0'dr_0'}
{(r_0 + \xi_r - r_0' - \xi_r')}.
\label{gs20}
\ee
We can re-write equation~(\ref{gs20}) in terms of the parameter, $q,$ with
\be
q= \frac{A(r_0) - A(r_0')}{r_0 - r_0'}. 
\label{q}
\ee
Then we find
\be
g_{D}(r_0) =  \frac{1}{2\pi} \int_0^{2\pi}\cos(\phi_0) \Delta  \left({{\partial
\psi_{SG}} \over {\partial r}}\right) d\phi_0
\ee
can be expressed as
\be
g_{D} = \frac{2G}{{\bar r}}\ \int_{r_1}^{r_2} \frac{I(q)}{q}
\Sigma(r_0')r_0'\ \frac{A(r_0) - A(r_0')}{(r_0 - 
r_0')^2} dr_0'
\label{gs2}
\ee
where:
\be
I(q) = \frac{1}{2\pi} \int_0^{2\pi} \frac{cos(\phi)}{ 1 - q\cos{\phi}} d\phi
= \frac{1}{q\sqrt{(1 - q^2 } }\ \left(1 - \sqrt{(1 - q^2 } \right) 
\ee
 Note that the integral in equation~(\ref{gs2}) being dealt with the principal value prescription
requires $\Sigma$ to vanish smoothly at edges  in order to avoid a
singularity. It can be particularly complicated to evaluate in practice.

\subsection{The pattern frequency}

If we multiply both sides of   Eq. ((\ref{pr5}))  by  $r_0\Sigma(r_0)\Omega(r_0)$ and
integrate radially across the ring, assuming that, consistently with the conservation of
momentum and angular momentum, the  integrated collisional terms vanish, 
we obtain:
\begin{eqnarray}
&\int_{r_1}^{r_2} {\Omega(r_0')}\ (\Omega_P -  \dot{\varpi}(r_0'))\
\Sigma(r'_0) r'_0A(r'_0) dr'_0 = \nonumber \\
&\hspace{-1cm} \left({2G}/{\bar r}\right)\int_{r_1}^{r_2} \int_{r_1}^{r_2} \left(I(q)/q\right)\Sigma(r'_0) \Sigma(r_0)
\left((A(r_0) - A(r_0'))/(r_0 - r_0')^2\right) r'_0 r_0 dr_0 dr'_0.   
\label{OmegaP1}
\end{eqnarray}
In addition the integral on the right hand side can readily be seen to 
vanish identically,  therefore we can calculate the value of the
pattern frequency,  $\Omega_P$, as:
\be
\Omega_P = \frac{\int_{r_1}^{r_2} \Omega(r'_0)\  \dot{\varpi}(r'_0)\
\Sigma(r'_0) A(r'_0) r'_0 dr'_0 }{   
\int_{r_1}^{r_2} {\Omega}(r'_0)\ A(r'_0)  \Sigma(r'_0) r'_0  dr'_0 },
\label{OmegaP}
\ee
We remark that for a thin ring $r_0'$ can be approximated as being constant in the above integrals
and thus may be cancelled out.

\subsection{The value of q }\label{qval}

In the linear regime: $q<<1$, $2I(q)/q\rightarrow 1$, so then
 multiplying  both sides of   Eq. ((\ref{gs2}))  by  $r_0\Sigma(r_0)\Omega(r_0)A(r_0)$ and
integrating  radially across the ring
we obtain

\begin{eqnarray} \label{q1} 
&\hspace{-8cm} \int_{r_1}^{r_2}
g_D\ \Sigma_0(r_0)\ A(r_0) r_0dr_0 =\nonumber \\
 &\hspace{-6mm}\left({G/2}\right)\int_{r_1}^{r_2}
\int_{r_1}^{r_2} \Sigma_0(r_0)\ \Sigma_0(r_0')\ \left(
\left({A(r_0)-A(r_0')})/({r_0 - r_0'} \right)\right)^2r_0 r_0' dr_0 dr_0' > 0
 \end{eqnarray}
On the other hand, from Eq.~(\ref{pr5}) we can write:
\be
\label{q2}
\int_{r_1}^{r_2}\ (\Omega_P -  \dot{\varpi} )\Omega(r_0)  A(r_0)^2 \Sigma_0(r_0)r_0
dr_0
= \int_{r_1}^{r_2}\ g_{ext} \Sigma_0(r_0)\ A(r_0)r_0 dr_0
\ee
Thus, if the  effect of collisions can be neglected in comparison to  that due to 
self-gravity,  it is verified that:
\be
\label{q3}
\int_{r_1}^{r_2}\ (\Omega_P -  \dot{\varpi} )\Omega(r_0)A(r_0)^2\ \Sigma_0(r_0)r_0
dr_0
> 0
\ee
We remark that in the linear regime the pattern speed, $\Omega_P$ may be regarded as an
eigenvalue associated with a normal mode determined by Eq.~(\ref{pr5}).
For a mode  which is such that   $A(r_0)$ does not change sign, as in
the case of interest here,
it follows directly from equation (\ref{OmegaP})
that $\Omega_P$ must equal the local precession
frequency,  $\dot{\varpi}$ at some intermediate point in the ring,
with radius  $\bar{r_0}$ say,  thus   $\Omega_P = \dot{\varpi}(\bar{r_0})$.
Thus with  the help of (\ref{OmegaP}) and  (\ref{q3}) , we can verify that 
\be
\label{q4}
\int_{r_1}^{r_2}\ (\Omega_P -  \dot{\varpi})\ A(r_0)\Omega(r_0)
(A(r_0)-A(\bar{r_0}))\ \Sigma_0(r_0)r_0 dr_0
> 0,
\ee
which with the use of the intermediate value theorem can be written as
\be
\label{q5}
\int_{r_1}^{r_2}\ (\Omega_P -  \dot{\varpi})\  A(r_0)
{\tilde q}(r_0 - \bar{r_0})\Omega(r_0) \Sigma_0(r_0)r_0dr_0
> 0,
\ee
where ${\tilde q}$ is a value of $q$ intermediate between $r_0$ and ${\bar{ r_0}}.$

For a thin ring we may perform a first order Taylor expansion and set 
 $ \Omega_P -  \dot{\varpi}(r_0) = -(d  \dot{\varpi}/dr_0)(r_0 - \bar{r_0}),$
 where the derivative is evaluated at $\bar{r_0}.$
Inspection of the integrand in (\ref{q5}) implies that if $A(r_0)$ and $q$ do not change sign we must have
 \be  
 -(d  \dot{\varpi}/dr_0)\ \times qA(r_0) > 0.\ee
As the precession frequency decreases outwards,  the eccentricity gradient 
is necessarily positive in any thin ring where self-gravity is the main
mechanism
that maintains apse alignment  and the gradient  does not change sign (see also Goldreich \& Tremaine 1979,
Borderies et al. 1983).

To estimate the value of $q$ in a narrow-eccentric ring,
in the linear regime where $q<<1$, $2I(q)/q \approx 1$  when
 all perturbations other than self-gravity are neglected,  we begin by noting that from
Eq.~(\ref{pr5}), we can estimate
\be
|{\bar \Delta}  \dot{\varpi}|A = \frac{GM_r|{\bar \Delta} A |}{2\pi {\bar r}\Omega  ({\bar \Delta}
r)^2}
\label{q6}
\ee
Here ${\bar \Delta} Q$ for some quantity $Q$
estimates its variation across the ring, thus
 $|{\bar \Delta } \dot{\varpi}|$ gives the magnitude of the difference
between
the free particle precession frequencies at the ring edges. Hence making the identification $e=A/{\bar r}$
we find that
\be |q| \sim {\bar r}{ |{\bar \Delta} e |\over |{\bar \Delta }r|} = {2\pi e \Omega {\bar r}^2
{\bar \Delta} r{\bar \Delta}
 \dot{\varpi} \over GM_r}
\label{valueq}
\ee
Hence, the magnitude of $q$ is related to  the mass, size and
eccentricity of the ring and the value of the second degree harmonic of
the gravitational potential of the central object, $J_2$.

\subsection{Conservation of radial action} 
\label{RA}

\noindent Following  (PM05) we define 
\be 
\label{Ir}
I_r = \int \Sigma_0 \Omega \left( {\partial \xi_r\over \partial
\theta_0}\right)^2 r_0 dr_0 d\theta_0 ,\ee where the integral is taken 
radially and azimuthally over domain of the ring. For $m=1$  radial
displacements of the form, $\xi_r = A(r_0)\cos(\phi_0) \equiv   e{\bar r} \cos(\phi_0),$  
as adopted above, for a slender ring  this integral corresponds
to the standard radial action, $\sqrt{GM_0a}
(1-\sqrt{1-e^2}),$ expanded to first order in $e^2.$
Here the semi-major axis, $a = {\bar r}.$
 
An expression for 
a conservation law for the radial action
associated with the $m=1$ mode can be found  by finding 
an expression for the time derivative of
Eq. (\ref{Ir}) as was done in  (PM05). To do this we return to the inertial frame 
and adopt $r_0,\theta_0,$ and $t$ as independent variables
while retaining the possibility of general time dependence
that is assumed to be slow as  compared  to the orbital time scale.
Thus we write
\be
\frac{d}{dt} \rightarrow \frac{\partial}{\partial t}+  \Omega \frac {\partial}{\partial \theta_0}.
\ee
and we recall that in linear theory, $\Omega$ can be replaced by the unperturbed value.
Also we have
\be
\frac{d^2}{dt^2} \rightarrow   \frac {\partial^2}{\partial t^2}+ 
2\Omega \frac{\partial^2}{\partial t\theta_0}
+ \Omega^2 \frac {\partial^2}{\partial \theta_0^2}.
\ee
The assumption of slow evolution enables us to neglect the first term in the above
and the assumption of an $m=1$ harmonic dependence on, $\theta_0,$ then allows us to write
\be
\frac{d^2}{dt^2} \rightarrow   \frac {\partial^2}{\partial t^2}+  
2\Omega \frac{\partial^2}{\partial t\theta_0}
- \Omega^2.
\ee
Using the above Equation (\ref{pr2}) take the form
\be
2\Omega \frac{\partial^2\xi_r}{\partial t\theta_0} + \xi_r (\kappa^2-\Omega^2)
= f_{r} - \Delta \left({\partial \psi '\over \partial r} \right)  + 2 \Omega {\cal Q}_\theta
\label{pr2A}\ee

%
On multiplying (\ref{pr2A}) by $\partial\xi_r/\partial\theta_0$
and integrating over  the mass of the ring, after performing an integration by parts
and  with the help of Equation (\ref{pth0}) we obtain
\begin{eqnarray}
&d\left(\int ^{r_2}_{r_1} \Sigma(r_0)\Omega 
\left(\partial\xi_r/\partial\theta_0\right)^2 r_0 d\theta_0 dr_0\right)/dt= dI_r/dt \nonumber \\
&= \int ^{r_2}_{r_1} \Sigma(r_0)\left(\partial\xi_r/\partial\theta_0\right)
\left( f_{r} - \Delta \left(\partial \psi '/ \partial r \right)  
\right)r_0 d\theta_0dr_0+\nonumber \\
&\int ^{r_2}_{r_1} \Sigma(r_0)\left(\partial\xi_{\theta}/\partial\theta_0\right)
\left( f_{\theta} - \Delta \left(1/r \left(\partial \psi '/ \partial \theta \right)  
\right)\right)r_0 d\theta_0dr_0.
\label{pr2A1}\end{eqnarray}
We remark that in obtaining Equation (\ref{pr2A1}) we have neglected the time derivative of 
${\cal Q}_{\theta}$ when making  use of  Equation (\ref{pth0})  and finally used the relation
$\partial\xi_{\theta} /\partial\theta_0 = -2\xi_r$ which applies to free epicyclic motion
in order to obtain this term in the second integral on the right hand side. 

For a ring in which a steady eccentricity is maintained,
noting that there is no net contribution from the ring's  self-gravity (the terms involving $\psi'$),  
 the secular rate of change of $I_r$
 due to satellite forcing must balance that due to the effect of
collisions  acting on the $m=1$ perturbation. 
 From (\ref{pr2A1}) this is seen to be equal to the ratio of the rate of energy dissipation 
associated with them that is induced by the $m=1$ perturbation  to the local orbital frequency.
In such a case that is the
condition that on average, $I_r,$ remains constant. 

As  there is no net contribution from self-gravity when integrated
over the mass of the ring,  any dissipation arising from internal
friction can only be compensated by the action of external forces that produce  satellite
torques. The exact multiple of a satellite torque that acts on the ring so
as to maintain the $m=1$ mode against energy dissipation  is considered in the next Sections.  

\subsection{Resonant satellite torques}\label{RST}
\label{SATT}

 Consider a satellite in circular orbit  with angular velocity $\omega$.  We suppose  that there is a second order
Lindblad  resonance at some point in the ring where
\be (m\pm 1)(\Omega -\Omega_P) = m(\omega -\Omega_P)
 \mp\kappa \label{rescond}\ee
for azimuthal mode number, $m.$
The upper/lower signs respectively corresponding to an interior/exterior satellite.
Note that this condition is expressed in terms of the orbital rotation frequencies as
seen in a frame rotating with the ring precession frequency. 
If we neglect orbital precession by setting $\Omega_P=0$ and $\kappa = \Omega,$
this gives the condition $(m\pm 2)\Omega= m\omega .$

For large $m,$ this corresponds to the satellite being a distance $\Delta_s \sim 4{\bar r}/(3m)$
from the nearest ring edge. 
If $m$ is large, as we shall assume, the satellite may have both first and second order Lindblad resonances
within the ring. The first order resonances may be associated with ring confinement,
leading to a shepherding role for the satellite,
the second order resonances being associated with eccentricity driving.
In order for there to be a second order resonance but no first order resonance
the ring must be sufficiently thin, such that ${\bar \Delta}r < \sim  4{\bar r}/(3m^2).$
When the situation is marginal a single satellite may have a first order resonance at the ring edge
providing confinement together with second order resonances at the edge and in the ring interior
which can drive eccentricity.

Returning to the case of a second order resonance expressed by
(\ref{rescond}) the contribution of the terms involving a satellite
potential in the integrals on the right-hand side of Eq.~(\ref{pr2A1})
are directly related to the resonant torque generated by the $m\pm1$
forcing between the satellite and ring as occurs in Eq.'s~(\ref{pr})
and~(\ref{pth0}) through its contribution to the forcing gradient of
potential (see PM05). 

The forcing amplitude is proportional to both the satellite potential and the
ring eccentricity. We neglect 
additional forcing terms that can arise
from the components of $( f_r, f_{\theta})$ with azimuthal mode number $m \pm
1$, and we assume that the ring-satellite torque is then
produced by the direct forcing of the unperturbed background ring 
a procedure that  is assumed adequate to provide the total torque.

The total torque, ${\dot J}_{m\pm1}.$  induced by a satellite on the eccentric ring
 can be estimated as in Goldreich \& Tremaine (1978).
An expression within the framework of the Lagrangian displacement is
given in PM05.
If the resonance occurs with ring material with angular velocity $\Omega,$ 
the disk is forced by a disturbance with a
definite pattern speed, $\Omega_{PP} =\Omega  \pm \kappa /(m\pm 1),$  and the rate of change of ring orbital
energy  is related to the rate of change of ring angular momentum  by 
\be  {\dot E}_{m\pm1} =
\Omega_{PP} {\dot J}_{m\pm1}, 
\label{balance}
\ee 
which expresses the well known result that the ratio of 
energy to angular momentum exchanged is $\Omega_{PP}$  
 \citep[see also][]{FriedmanandSchutz.1978} .
We also have the contribution to ${\dot I_r}$ given by
\be
{\dot I_r} = { {\dot E}_{m\pm1} \over
\Omega} - {\dot J}_{m\pm1} =\pm {\dot J}_{m\pm1}\kappa/((m\pm 1)\Omega).\label{actdot}
\ee 
Note that ${\dot J}_{m\pm1}$ is respectively positive/negative for an interior/exterior satellite
so that the contribution to ${\dot I}_r$ is always positive resulting in ring  eccentricity growth.
Of course the contributions from all satellites and relevant resonances must be incorporated. 
These positive contributions to ${\dot I_r}$ have to be balanced by effects due to collisions.

\subsection{Estimate of the magnitude of an eccentricity driving satellite torque}\label{RST1}

We use the balance Equation~(\ref{pr2A1}) to estimate the masses of
 the putative shepherd satellites of (10199) Chariklo's ring system as
 a function of their location with respect to the ring. Since here we
 are only interested in an order-of-magnitude estimation, we shall
 only consider an interior satellite in a circular orbit with an
 associated exterior second order resonance. The corresponding
 calculation for an exterior satellite with an inner resonance is
 entirely analogous.

We estimate the satellite torque resulting from the $(m+1)$ forcing
described in Section \ref{SATT} from results of \citep{GoldreichandPorco1987} who give

\be {\dot J}_{m+1}= 3 e^2 m^4 \ \Sigma_0\ \left(\frac{M_{sat}}{M_{0}}
\right)^2 \Omega_0^2 {\bar r}^4 \label{torquem+1} \ee where we recall
that ${\bar r} $ is the mean distance from the ring to the central
body, $\Sigma_0$ is the unperturbed surface density of the ring, 
the eccentricity of the ring, $e,$ and $\Omega_0$ here being evaluated
at the location of the resonance,$M_{sat}$ is the mass of the
perturbing satellite and ${M_{0}}$ the mass of the central body.  The
contribution to the rate of change of radial action may then be found
from (\ref{actdot}) with the choice of upper sign.

\section{The energy dissipation rate in an asteroidal ring and the masses of
 perturbing  satellites}
\label{CH_ring}
  
To estimate the energy dissipated per unit mass in the ring, we
envisage that local velocity differences which might arise from
streamline intersection produced by differential precession or from
the action of external satellite torques, are counteracted by forces
that arise through collisions. It has been suggested that these could
occur as a consequence of close-packing that ring-particles experience
when they go through pericenter \citep{DermottandMurray1980} or an
increased velocity dispersion produced through the ring satellite
interaction \citep{ChiangGoldreich2000}. In order to estimate the rate
of energy loss due to these effects we adopt the following heuristic
approach.  We introduce a velocity scale through
$u_c =  e {\bar r} {\bar \Delta}  \dot{\varpi},$
where $\Delta \dot{\varpi}$ is the difference between the precession
frequency induced by the oblateness of the central object at the edges
and $e$ is the eccentricity. This is roughly the radial velocity
difference between the edges at pericentre that is induced by
differential precession on an orbital time scale after initial
alinemment.

We write the energy dissipation rate as, 
${\dot E}_{dissip}^{CP}$,  where: (see also  PM05)
\be 
{{ {\dot E}_{dissip}^{CP} }\over{\Omega_0}} = 2 \pi \beta \Sigma_0  {\bar \Delta} r {\bar r} u_c^2, 
\label{ED1}
\ee
where we recall that ${\bar \Delta} r$ is the width of the ring
and $\beta$, a  scaling factor  which in addition to scaling the velocity could  include  effects of inelasticity
through a restitution coefficient.
We remark that the above rough estimate,  with $\beta$ of order unity, is expected to be of the correct order of magnitude
if $u_c$ gives the correct velocity scale for dissipation on an orbital time scale.
We remark that the postulated enhanced collisional effects that may occur very close to sharp edges \citep{ChiangGoldreich2000} may 
be incorporated through an appropriate choice of $\beta.$
  
 In the case of the $\epsilon$-ring of
Uranus, it has been shown that, when collisional  effects,
as estimated above,   with $\beta \approx 1$,  are
assumed as the main cause of dissipation that is balanced by the rate
of increase of the ring radial action driven by a perturbing
satellite, Eq.~(\ref{actdot}) together with (\ref {torquem+1}) yields
excellent agreement with the required ratio of energy input rate to
angular velocity found from (\ref{ED1}) when the mass and orbit of the
shepherd satellite Cordelia (PM05) are adopted.

\subsection{ Viscous dissipation in the background flow}\label{viscd}
However, apart from effects arising from a non zero eccentricity,
 there is a rate of energy dissipation associated with an effective  viscosity acting on 
 the background Keplerian flow,  ${\dot E}_{dissip}^{BKF}$, which is given by
\citep{LyndenBellandPringle1974} 
\be 
 {\dot E}_{dissip}^{BKF}   = \frac{9}{4} \nu \Omega^2 M_r,
\label{ED2}
\ee
where $M_r$  is the mass of the ring. In order to make the estimates below
we  evaluate $\Omega$ and $\nu$  at the resonance  radius  $r=r_0.$
Thus $\Omega \rightarrow \Omega_0,$ and $\nu \rightarrow \nu_0.$
In addition we estimate the local magnitude of the kinematic viscosity 
as $\nu~=~H^2~\Omega_0,$, where $H$ is the associated semi-thickness
$H=\Sigma_0/(2\rho_0)$ with $\rho_0$ being the ring density at $r=r_0.$

We suppose that eccentricity driving occurs through the operation
of a second order resonance with a single interior satellite and this
is balanced by energy dissipation through impulsive collisions 
for the most part near the ring edges.
 This implies that the satellite radial action input
rate given by Eq.~(\ref{actdot}) obtained with the help of (\ref
{torquem+1}) balances with the ratio of the energy dissipation rate to
angular velocity found from (\ref{ED1}).  This condition enables the
mass of the satellite to be expressed in units of the mass of the
central object as 

\be M_{sat}^{CP} = \left( {2 \pi \beta {\bar \Delta} r (m+1)}
\over {3 \Omega_0^2{\bar r}} \right)^{1/2}\ \left( {{\bar \Delta}
 \dot{\varpi}}  \over {m^2} \right) M_{0}, \\
\label{MASS_SAT1} 
\ee

For comparison we give the satellite mass under the assumption that
the rate of increase of radial action induced by  the satellite is
balanced by the ratio of the background dissipation rate to angular
velocity in the form 
\be M_{sat}^{BKF} = \left( {1.5\ \pi {\bar
\Delta} r \nu(m+1)} \over { {m^4} e^2 \Omega_0{\bar r}^3 }
\right)^{1/2}\ M_{0}, \\ \label{MASS_SAT2} \ee
This will exceed $M_{sat}^{CP}$ when the background viscous
dissipation exceeds that arising from  enhanced collisions arising
from satellite perturbations and close packing. It is important to
note that this is {\it not} a condition for the satellite torques to
prevent viscous spreading. In that case the satellite torque must
balance the viscous outward angular momentum flow rate 
\citep[e.g.]{GoldreichandTremaine1978} 

\begin{figure}
\begin{center}
\includegraphics[width=6.5cm,angle=270]{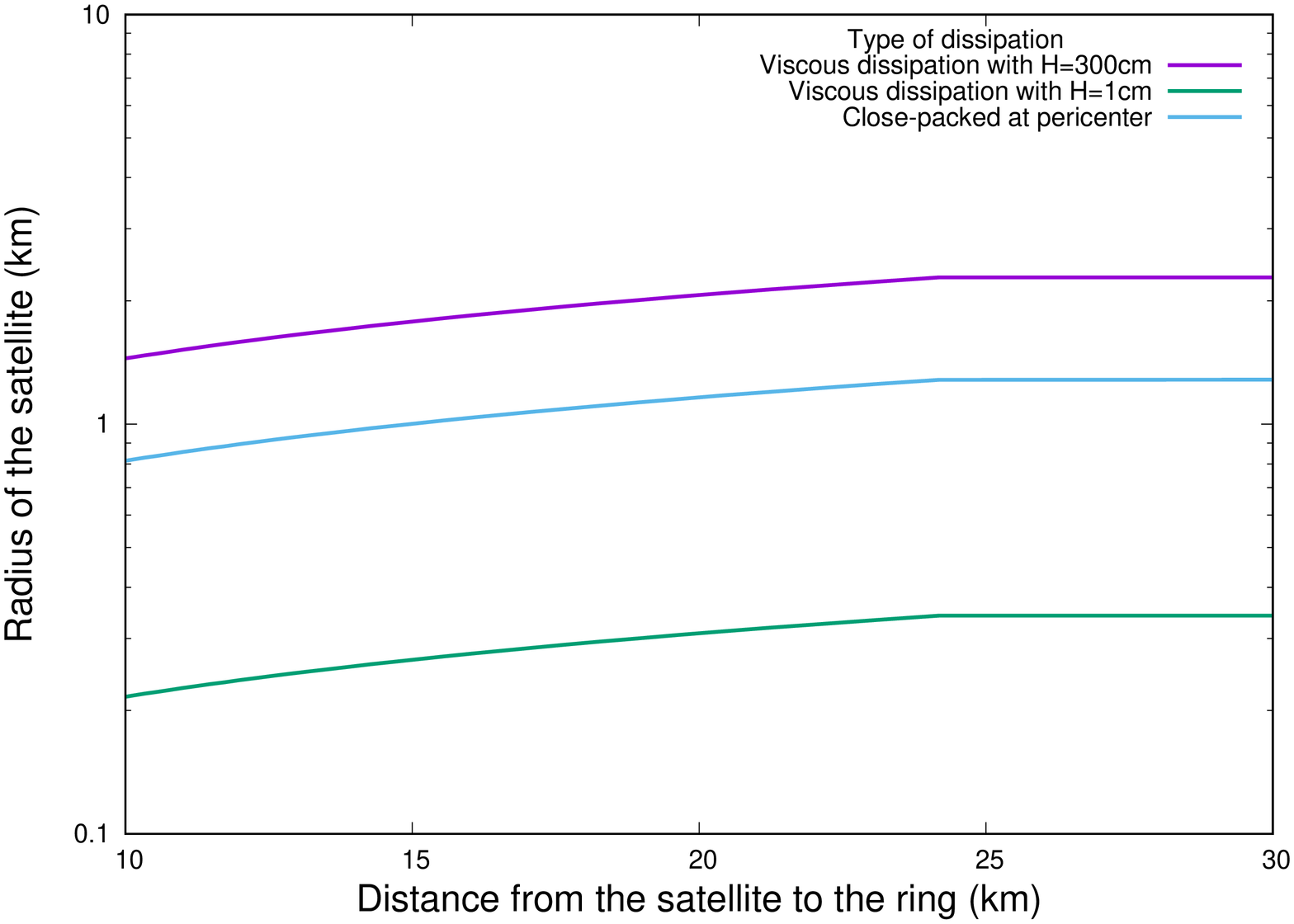}
\includegraphics[width=6.5cm,angle=270]{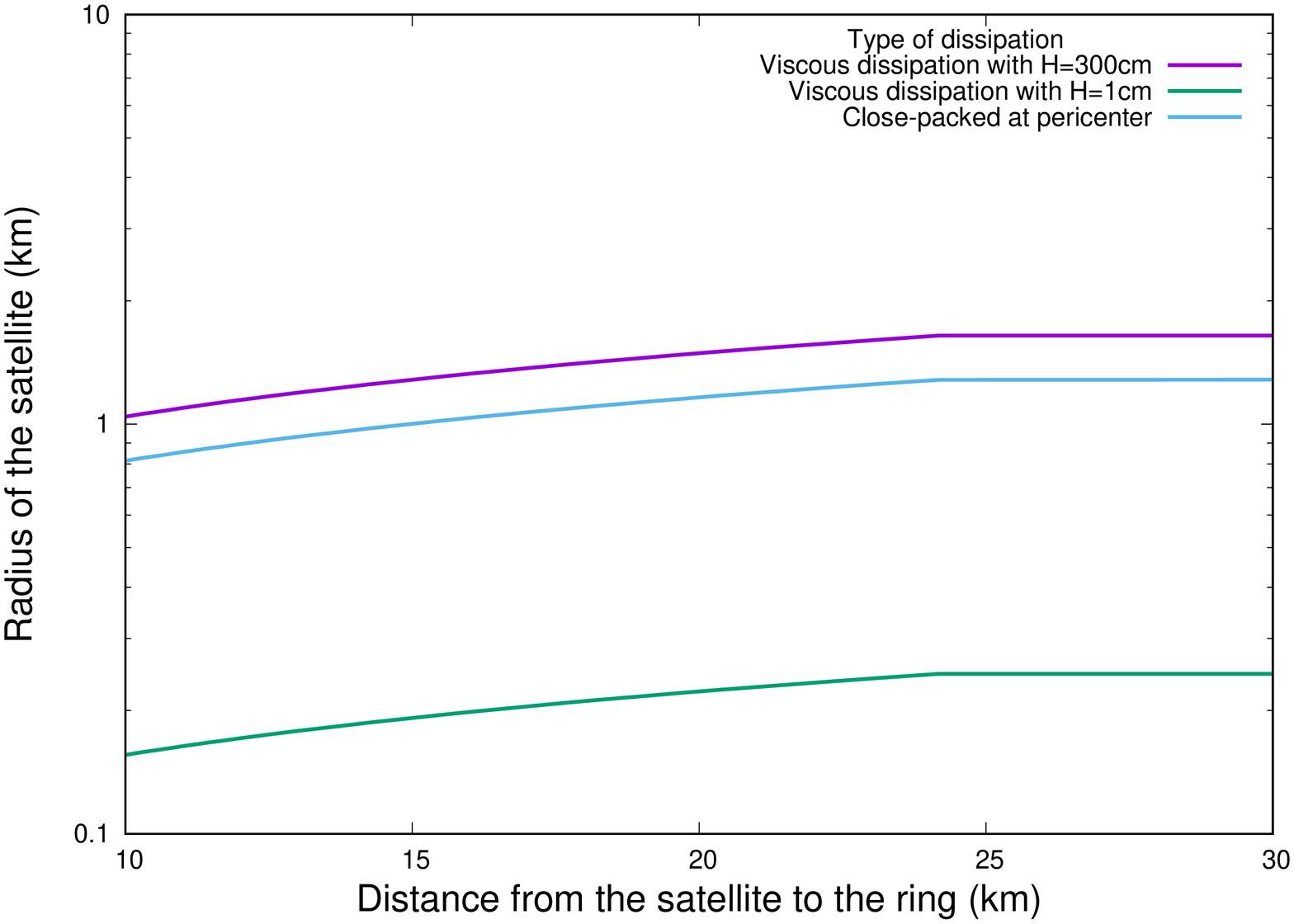}
\caption{Mass estimates for a perturbing  satellite 
which is able to balance the effect of two different types of rate of
ring energy dissipation given in the text are $M_{sat}^{CP} $ and
$M_{sat}^{BKF}$. For clarity we actually plot the values of their
corresponding radii as a function of their distance from the ring,
assuming them to be spherical and of uniform density $1 gcm^{-3}$.
When the dissipation occurred through enhanced collisions, the values
plotted are for $\beta =1.$ 
 The results plotted in the
upper panel are for the ring eccentricity $e =0.003$ while the lower
panel is for $e =0.008.$  A value of $J_2=0.13$ was assumed in all
plots.}
\label{fig:MASS_SAT}
\end{center}
\end{figure}


In Fig.~\ref{fig:MASS_SAT} we plot radius estimates
for an interior  perturbing satellite of (10199) Chariklo's ring system, as a
function of the distance from the satellite to the
ring, which can be approximated as 
\be \Delta_s ={\bar r} -  r_{Satellite}  = {\bar r}
\left(1- \left(m/(m+2) \right)^{2/3}  \right),\ee
which in turn can be used to express $m$ in terms of $\Delta_s$ in the form
\be m = \frac{2}{\left( 1- \Delta_s /{\bar r}\right)^{-3/2} -  1}.
\ee

The parameters of the ring system used to produce the plots are given 
Section~\ref{sec.Ch}.  The satellite resonance and enhanced collisions 
are assumed to be in the CR1 ringlet  with $\beta =1$. Two values of the eccentricity
of the ringlet have been adopted in equation (\ref{ED1}) when
evaluating the energy dissipation rate as a result of satellite forcing  and/or close packing,
 $e= 0.003$ for the calculation displayed in the upper
panel and $e= 0.008$ for the calculation displayed in the lower panel.
The energy dissipation rate due to viscosity was evaluated for the
semi-thicknesses, $H=300cm,$ and $H=1cm,$ respectively corresponding
to estimated values of $\Sigma_0 = 300 g cm^{-2}$, $\Sigma_0 = 1 g
cm^{-2}$ for an assumed $\rho_0 = 0.5 g cm^{-3}.$ The differential
precession rate across the ring, ${\bar \Delta} \dot{\varpi},$ was
taken to be $-7.93\ 10^{-8} s^{-1}$, which corresponds to a value of
$J_2 = 0.13$ and the location and width of the CR1 ringlet were adopted.  Note that
in the case where dissipation arises through  satellite forcing and/or close packing, 
 the estimated radius of a satellite necessary to
counteract the decay is between $30 \%$ and $50 \%$ smaller than the
values plotted in Fig.~\ref{fig:MASS_SAT} for assumed values of
${\beta}$ between $0.1$ and $0.01$, and then its size is smaller than
that of a satellite needed to achieve confinement (BR14).  The energy 
produced by viscous dissipation is significantly larger than that
arising through enhanced collisional effects  when the assumed surface
density is $\Sigma_0 = 300 g cm^{-2}$ and it is similar when the
assumed surface densty is $\Sigma_0 = 1 g cm^{-2}$ .

 We may also estimate the potential importance of the 1:3 second
order resonance between the mean orbital frequency in a circular orbit
and the intrinsic rotation frequency of (10199) Chariklo. This is
close to the inner edge of CR1 and could act to balance dissipative
collisional effects arising from close packing were this resonance to
be inside the ring. \citet{SICARDYETAL2018} estimate that the non
spherical deformation of (10199) Chariklo could amount to a mass
anomaly of $10^{-5}M_0.$ This corresponds to an object of radius
$2.4km$ which is about a factor of two larger than the radii indicated
in Fig.\ref{fig:MASS_SAT}. However, the object has to be regarded as
more than an order of magnitude further away. The corresponding value
of $m =2,$ whereas for objects indicated in Fig.\ref{fig:MASS_SAT},
$m\sim 16.$ Use of equation (\ref{torquem+1}) for this virtual mass
indicates that if the resonance is indeed inside the ring the torque
input is about an order of magnitude smaller than for those considered
in Fig. \ref{fig:MASS_SAT}. However, significant effects could arise
if the mass anomaly was three times larger than that assumed. In that
case it could act together with closer satellites to maintain the ring
eccentricity.

%
%
%





\vspace{1cm}
\section{The condition of apse-alignment  across the ring and the 
surface density}
\label{aps-al}

\subsection{Probing the model of apse-alignment  for case of the $\epsilon$-ring of
Uranus}\label{probU}

On the left hand side of Eq.~(\ref{pr5}) we have the product of the
local rate of differential precession relative to an assumed
precession of the ring as a whole and its local  radial displacement,  
the latter being the  product of the  local radius and  eccentricity.  On the
right hand side of Eq.~(\ref{pr5}) we find contributions from the
agents that can balance this and enable the entire ring to undergo
uniform precession, i.e. ring self-gravity and inter-particle
interactions.  

For the cases treated here there is no satellite with an orbit such
that its orbital frequency is resonant with the pattern precession of
the ring, as in the case of the ``Titan'' ringlet of Saturn that was
considered in \citet{MelitaandPapaloizou2005}. Accordingly, to
calibrate the model we will apply it to the $\epsilon$-ring of Uranus,
where independent estimates of the eccentricity, the
eccentricity-gradient across the ring and the surface density can be
 made from  Voyager 2 PPS $\beta$  Persei and $\sigma$  Sagittarii  stellar occultation
measurements \citep{Grapsetal1995}.

Our goal is to verify that the independent estimate of the value of
the surface density leads to contributions from self-gravity that can
potentially counteract the differential precession induced by the
oblateness of Uranus, across the ring.  In addition we compute the
contribution from inter-particle forces in those locations where the
contribution from ring self-gravity is unable to exactly match
that required to balance the kinematic rate of differential
precession.  

\begin{figure}
\begin{center}
\includegraphics[width=10cm,angle=270]{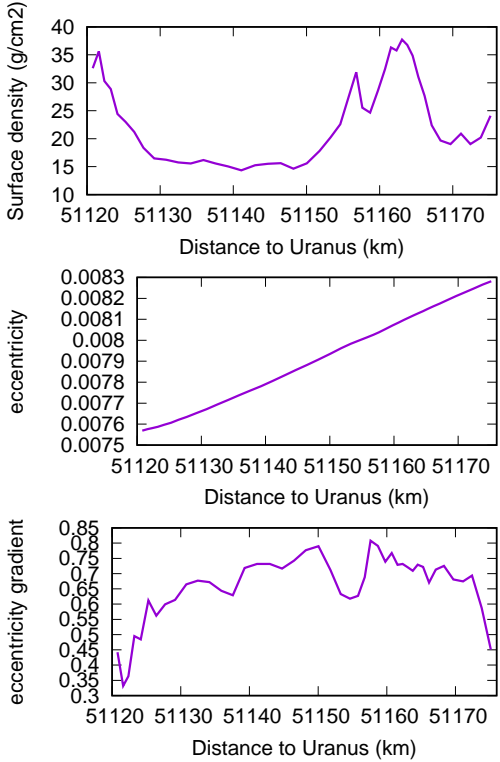}
\caption{Averaged surface-density, eccentricity and
eccentricity-gradient profiles for the $\epsilon$-ring of Uranus using
data obtained from \citep{Grapsetal1995}). 
}
\label{fig:graps}
\end{center}
\end{figure}

The input data from the $\epsilon$-ring, \citep{Grapsetal1995}, is
shown in  Fig.~\ref{fig:graps}. The surface density profile, $\Sigma(r)$, was estimated
from the optical-depth profile $\tau(r)$ from
$ \Sigma(r) = \frac{4}{3} \rho  a_p  \tau(r)$,
where  $\rho$ is the mean density of the   individual particles in the
ring, and $a_p$ is their radius.  If we assume that  $\rho=1 g\ cm^{-3}$
and  $ a_p  = 30\ cm, $
 we obtain an approximate
mean value of $<\Sigma(r)> = 30 g\ cm^{-2}$ (Fig. \ref{fig:graps}).
Unfortunately, there is not enough data to produce an accurate average
of the surface density 
over all azimuthal  angles. The input data that we use is obtained  from an average of
four different stellar occultation events. Some azimuthal structure may not be canceled
properly by our averaging procedure, 
which may explain some of   
the radial variation seen in the plot. Notice also that the
eccentricity gradient is remarkably constant across the ring, i.e. 
to a high degree of accuracy the
eccentricity increases linearly from one edge to the other.

In the Appendix we show how  
the self-gravity contribution, $g_D(r_0)$
is computed from the discrete data. This
data, \citep{Grapsetal1995}, contains information
from $41$ different locations across the ring,
which were re-sampled to obtain a total number of $n_{max}=450$ on a suitable equally spaced grid, since the
limit set in the Appendix to comply with the sampling theorem is about $n_{max}=400$.
This corresponds to a radial resolution of $\sim 130m.$
We recall that the edges of the ring are not resolved observationally or by the theoretical model at this level.

The azimuthally integrated inter-particle collisional contribution, $F_{cr} ~+~ 2\ F_{c\theta}$
is readily obtained from Eq.~(\ref{pr5})  as:
\be F_{cr} + 2\ F_{c\theta} =  {\Omega(r_0)}\ (\Omega_P -
 \dot{\varpi}(r_0))A(r_0) - g_D(r_0).
\label{eqFc}
\ee
We compute $A(r_0)$ from the available data as 
$$ A(r_0)= r_0\ e(r_0).$$
The quantity, $q$ is assumed  to be constant. To justify this we begin by  noting
that $q =r_0\ de/dr_0 + e.$ From the data plotted in  Figure~\ref{fig:dp}
below, we see that neglecting a fractional errors of order the $10\times$ 
the ring width to radius
ratio, the second term may be neglected and an approximately linear
variation of eccentricity implies that $q$ is approximately constant.
In this way, $q,$ becomes equivalent to the eccentricity gradient
introduced in Section \ref{sec.Intro}, being a situation that applies
to thin rings in general and we may write
\be
q \equiv {\bar r} \frac{de}{d r_0}.
\label{qq}
\ee
The precession frequency is given by
$$ {  \dot{\varpi}}(r_0) = \frac{3}{2}\ J_{2U}\ \Omega
\left(\frac{R_U}{r_0}\right)^2,
$$
where $J_{2U}$ and $R_U$ are  the second harmonic of the gravitational
potential and the physical radius
of Uranus respectively.  In Figure~\ref{fig:dp} we plot as a function of radial location in
the ring, the acceleration needed to  balance  the differential precession
due to the oblateness of Uranus, ${\Omega(r_0)}\ (\Omega_P -
{ \dot{\varpi}}(r_0))A(r_0)$, with $\Omega_P$  being given by equation (\ref{OmegaP}). 
Notice that  this quantity is positive in the inner
edge and negative in the outer edge, because the precession rate decreases
with distance. The acceleration due to self gravity  $ g_D(r_0)$ 
and the component arising from particle interactions, $F_{cr} + 2\ F_{c\theta}$, as obtained from
Eq.~(\ref{eqFc} ) are also plotted. Their sum balances the effect of differential
precession.
 
We note that close to the edges the collisional component is
significant and contributes in the same sense as the component induced
by the oblateness of the planet, while the self-gravity component that
together they balance is opposed and larger in magnitude than each of
them. 
\begin{figure}
\begin{center}
\includegraphics[width=6cm,angle=270]{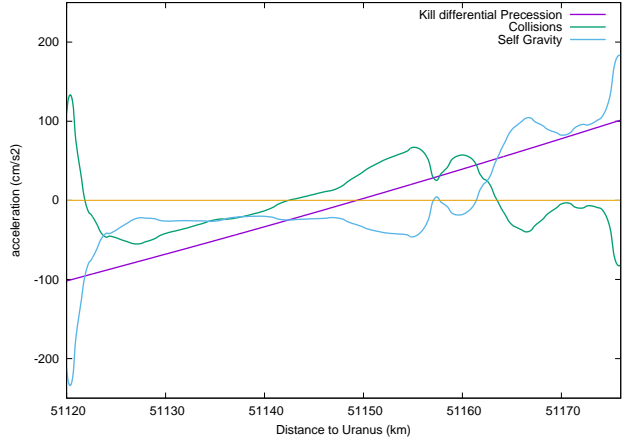}
\caption{Acceleration needed to cancel the differential
precession due to the oblateness of Uranus (magenta curve), the
acceleration due to self gravity (blue curve) and that due to
azumuthally integrated collisional effects (see equation (\ref{eqFc}))
(green curve), as a function of radial distance form Uranus. 
} 
\label{fig:dp}
\end{center}
\end{figure}

\subsection{The ring system of (10199) Chariklo}\label{Charring}

Our goal is to estimate the most plausible values of the 
assumed constant eccentricity
gradient and surface density for the CR1 and CR2 ringlets and, for
that purpose we assume that the real values of these parameters are
such that the integrated effect of
 collisions needed to  balance effects due to  
differential precession are  minimum values.
 We assume 2 possible extreme  values of $J_2$ for
(101999) Chariklo,  $J_2^a=0.28$, as taken from 
\citet{Leivaetal2017}  assuming that the surface of the body is
an equipotential determined by  the combined effects of centrifugal forces and  self gravity
  and $J_2^b=0.076$ as estimated by BR14 and
\citet{PanWu2016}. 

 We define the quantity, $F,$ as
\be
F =  \frac{1}{W} \int_{r_1}^{r_2} \left|\frac{F_{cr}(r') + 2\ F_{c\theta}(r')}{\Omega}\right| dr', 
\label{Feq}
\ee
where $r_1$ and $r_2$ are the inner and the outer radial boundaries
of the ringlet  respectively and $W=|r_2-r_1|$ is the width of the
ring. 
This quantity which we shall describe as an integrated
collisional term is an integrated collisional acceleration $/$ orbital
frequency.


\subsection{The CR1 ringlet}\label{CR1}

For the CR1 ringlet we assume a mean distance from the Centaur of
$394$km and an eccentricity at the inner edge of $0.003$. We compute 
the value of $F$ 
for assumed constant  values of $q$  (see equation (\ref{qq}))    
between  $0.01$ and $0.61$, which are compatible with the
presently available observations.  The upper bound of $q$ has been
chosen to be similar to that found for the planetary
eccentric ringlets.

The surface density profile
is assumed to be constant between the inner and the outer edge, 
and we explore values that range between $10 g cm^{-2}$ and  $610 g
cm^{-2}$. We note that the assumption of a uniform interior   surface
density qualitatively agrees with the observation of a 
uniform optical-depth profile (BR14
and B17). The top hat profile is justified on account of
the sharpness of the
rings being  observed to be much smaller than the resolution of the
occultation data (BR14 and B17). 
 The resolution of the grid used for numerical modelling (see the Appendix)
was approximately $20m$ and so the edge is not resolved at smaller scales.
We also assume that the eccentricity gradient is positive and constant
across the ringlet and therefore, the eccentricity increases linearly
between the inner and the outer edge. 


%

The values of surface density, eccentricity gradient that minimize the
collisional term $F$ are given in table \ref{tab:mins} for both
ringlets.

\begin{table}[ht]
\begin{tabular}{|l|ccc|ccc|} \hline
                 &   \multicolumn{3}{|c|}{CR1}       &
		 \multicolumn{3}{|c|}{CR2} \\ \hline
     &  $\Sigma$ ($g\ cm^{-2}$) & $q$    & $F$ ($cm\ s^{-1}$) &     $\Sigma$ ($g\ cm^{-2}$) &  $q$  & $F$ ($cm\ s^{-1}$) \\ \hline \hline
$J_2^{a}=0.28$   &         10.0 & 0.21  & 7.90       & 10.0 & 0.06 & 2.16 \\ 
                 &         10.0 & 0.29  & 13.37      & 22.2 & 0.03 & 3.88  \\ \hline \hline
$J_2^{b}=0.076$  &         10.0 & 0.01   & 7.45       & 10.0 & 0.01  & 0.86 \\
	         &         10.0 & 0.01   & 7.47      & 10.0 & 0.02 & 1.21 \\ \hline
\end{tabular}
\caption{Values of surface density, $\Sigma$ and eccentricity
gradient, $q$, that
minimise the collisional term, $F$, for ringlets CR1 and CR2 are given in the first entry under the
assumed values of  $J_2^a$ and $J_2^b.$  In each
case the value of the colllsional term $F$ obtained from equation
(\ref{Feq}) is given. The values given in the second entry correspond to a model for  which
the self gravity  contribution almost exactly balances the contribution of  differential precession 
in the middle section of the ring
 removing the need for a significant contribution from the collisional term there.
}
\label{tab:mins}
\end{table}
					 
\begin{figure}
\begin{center}
\includegraphics[width=6cm,angle=270]{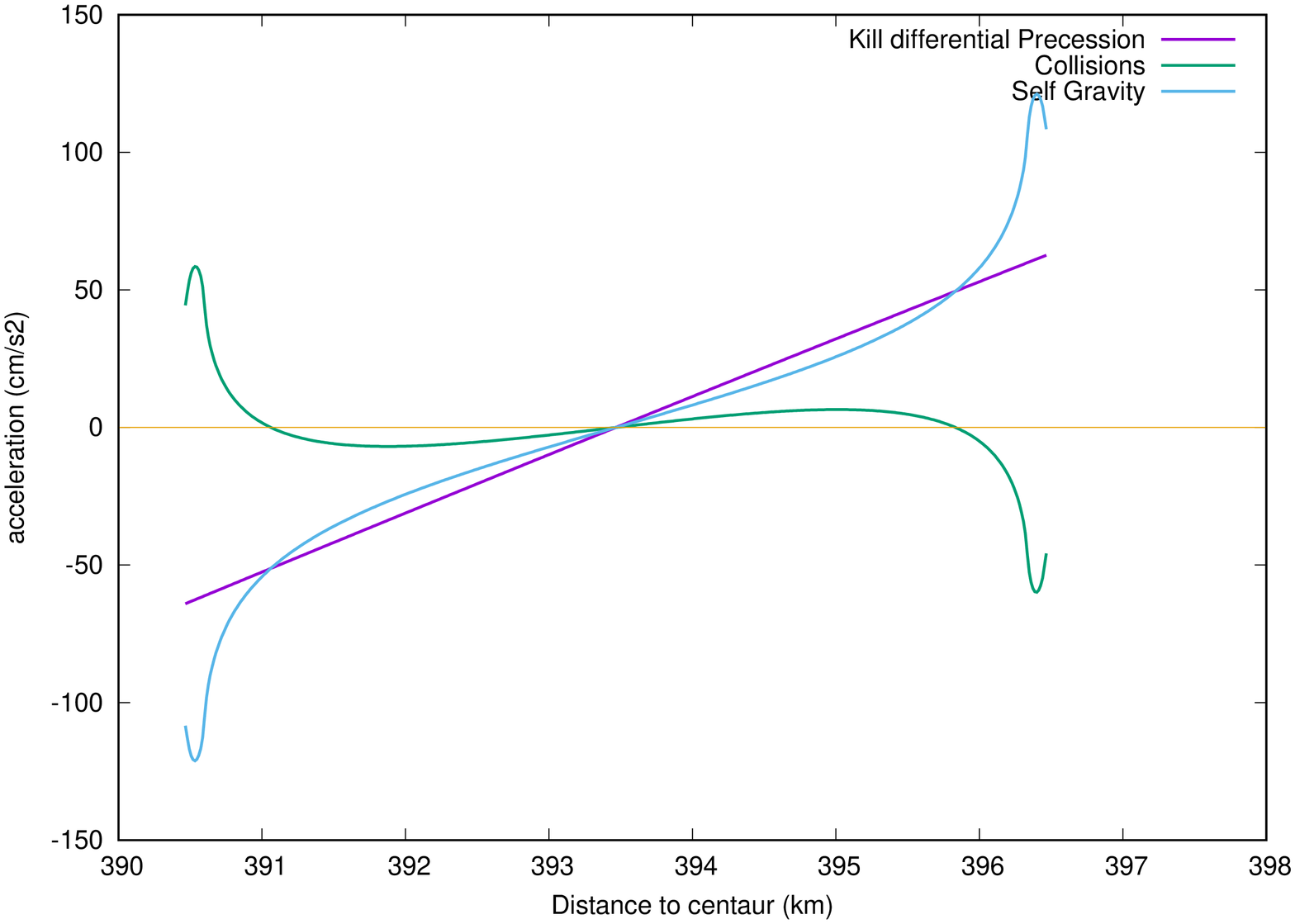}
\includegraphics[width=6cm,angle=270]{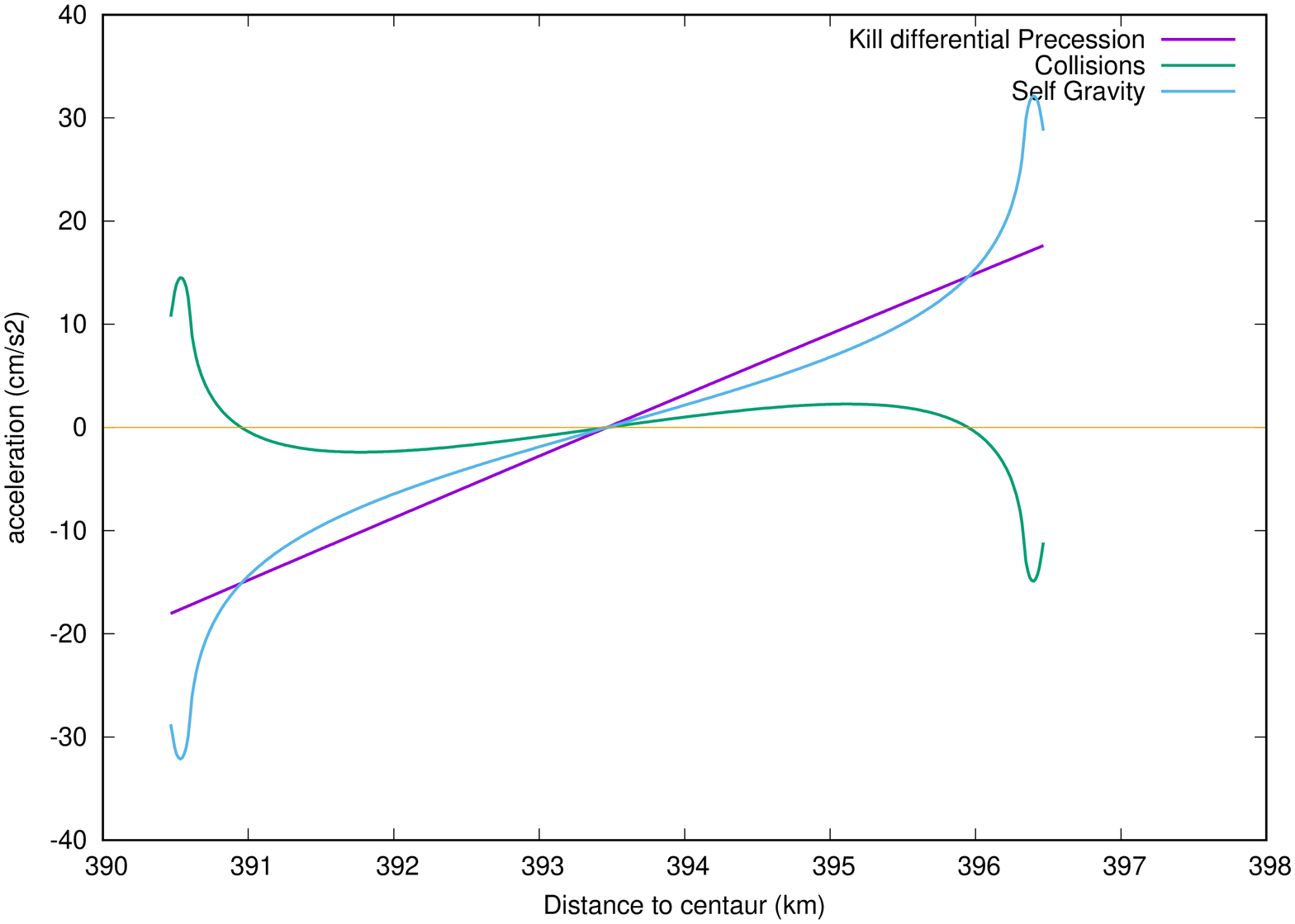}
\caption{ Above: The acceleration needed to cancel the differential precession due to
the oblateness of (10199) Chariklo (magenta curve), the acceleration
due to self gravity (blue curve) and that due to azimuthally
integrated collisional effects (see equation (\ref{eqFc})) (green
curve) for the CR1 ringlet model that minimises the integrated
collisional term, $F$, for a value of $J_2^a=0.28$. Below: Same as
above for a value of $J_2^b=0.076$.  For the values of  surface
density and eccentricity gradient used to produce each plot 
see Table \ref{tab:mins}.} 
\label{figCS1}
\end{center}
\end{figure}
\begin{figure}
\begin{center}
\includegraphics[width=6cm,angle=270]{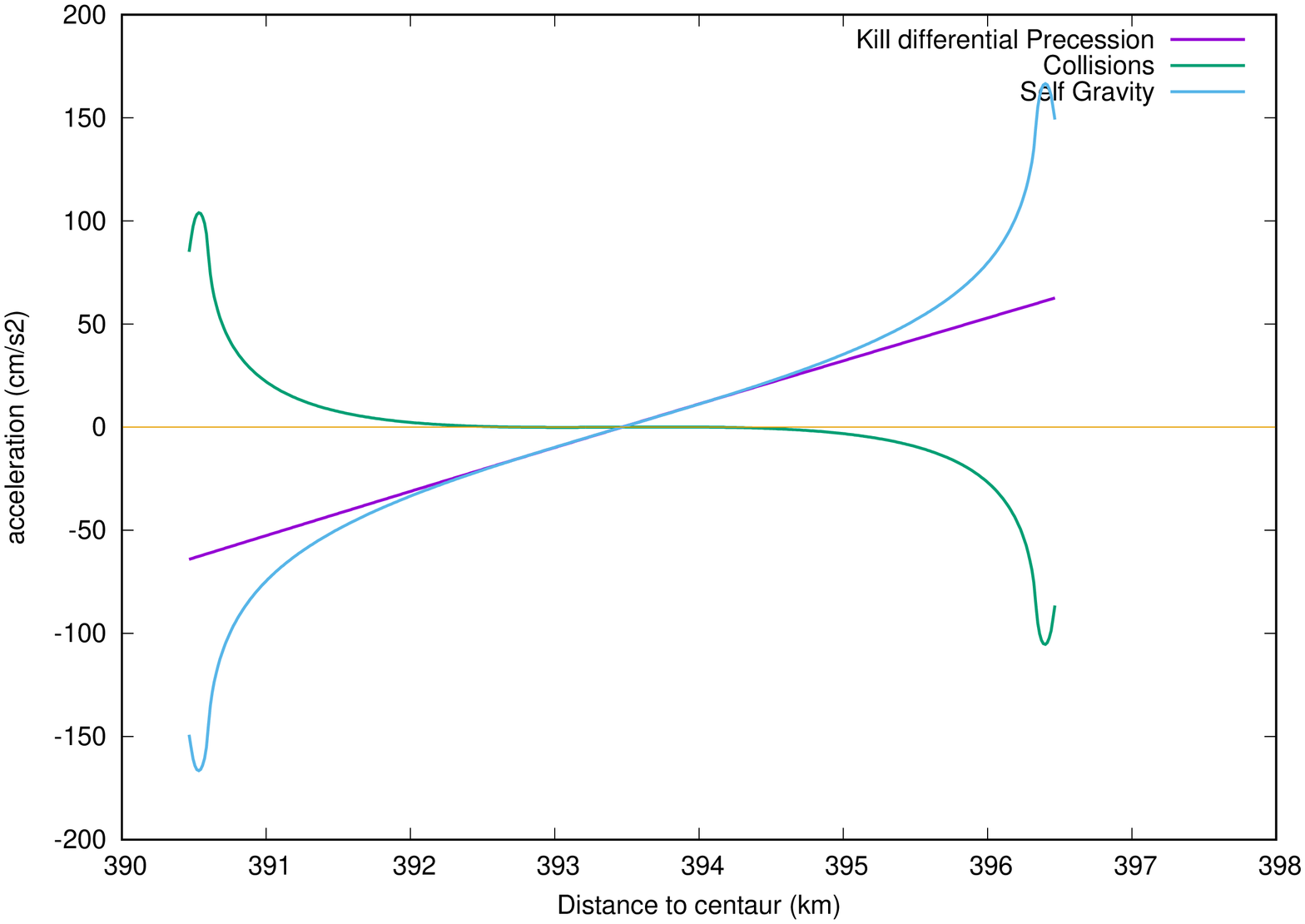}
\includegraphics[width=6cm,angle=270]{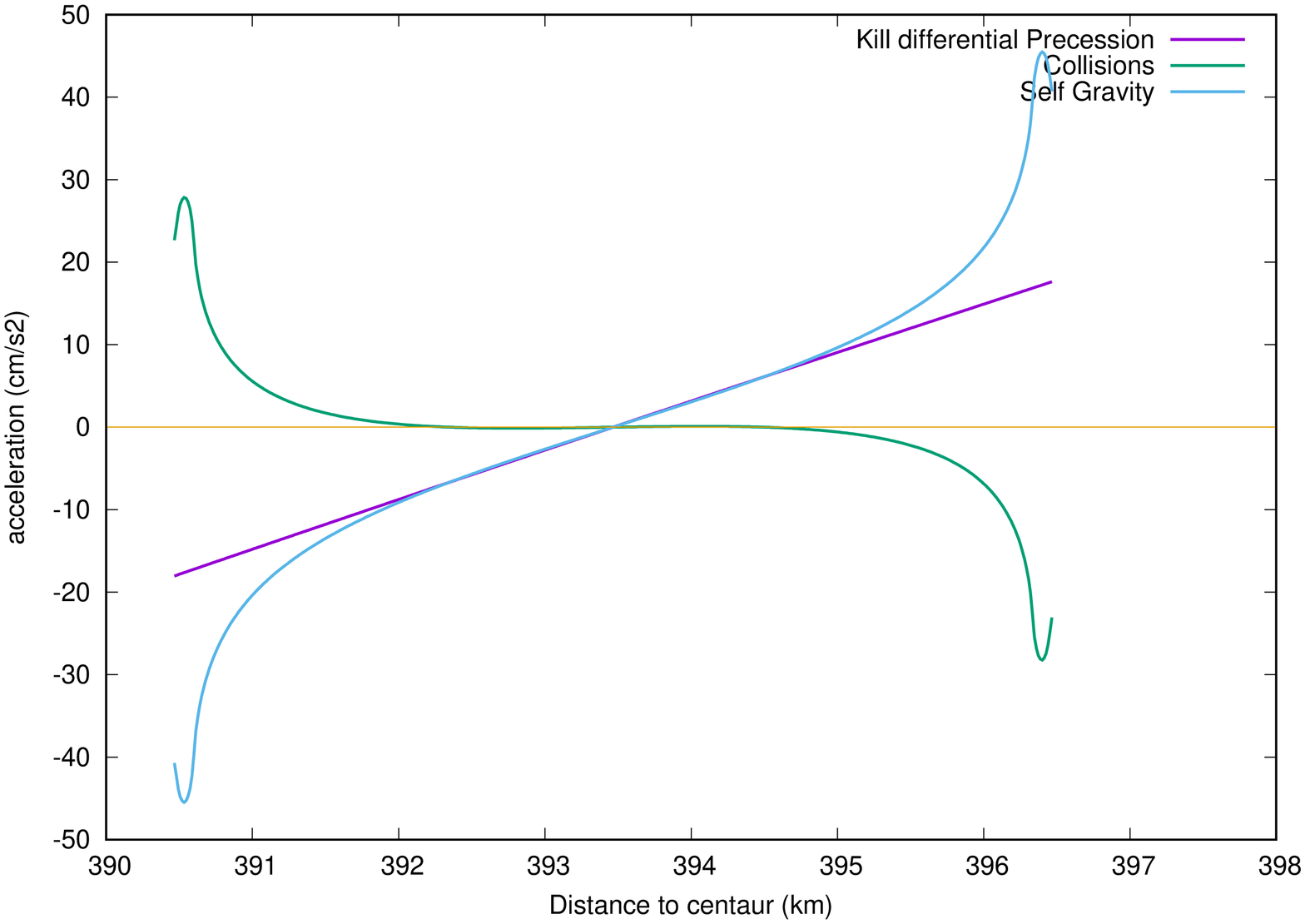}
\caption{
The acceleration needed to cancel the differential precession
due to the oblateness of (10199) Chariklo (magenta curve),
the acceleration due to self gravity (blue curve)
and that due to azimuthally integrated collisional effects (see
equation (\ref{eqFc}))  (green curve)
for the CR1 ringlet model that eliminates  the need for
the collisional term in the middle section of the ring, for a value
of $J_2^a=0.28$ (above) and $J_2^b=0.076$ (below).  {The values of
surface density and eccentricity gradient used to produce each plot
can be found in table \ref{tab:mins}}.}
\label{figCS11b}
\end{center}
\end{figure}

 In the case in which $J_2^{a}=0.28$ the integrated collisional
term, $F$, for the CR1 ringlet is minimised for a value of $q=0.21$,
and a value of the surface density of $\Sigma = 10 g cm^{-2}$.  The
radial distribution of accelerations for this case is illustrated in Figure
\ref{figCS1}.  For the model in which the slopes of the self-gravity
 contribution and the differential precession contribution  are equal at the
middle of the ringlet    removing the need of a
significant  contribution from the  collisional term there, the overall dissipation is about twice as
large as in the previous model for the same value of the surface
density and a similar  but larger value of eccentricity gradient. The
corresponding radial distribution of accelerations is illustrated in
Figure \ref{figCS11b}..



%
%
%
 
For a lower value of $J_2^b=0.076$ we find that the collisional
acceleration is minimised  for a smaller value of the eccentricity
gradient, $q=0.01$, and a
value of surface density,  $\Sigma = 10 g cm^{-2}$, as in teh
previous case.  The radial
distribution of
accelerations
for this case is illustrated in Figure \ref{figCS1}. For
$J_2^b=0.076$, the model 
in which 
the self-gravity compensates the differential
precession induced by the oblateness of the Centaur in the middle section  of
the ringlet occurs for a larger value of surface density, producing a slightly
larger collisional term $F$. The corresponding radial
distribution of
accelerations is illustrated in Figure \ref{figCS11b}.

For the models where the integrated collisional term is minimised and   self-gravity {\it approximately} compensates the differential
precession induced by the oblateness of the Centaur in the middle of
the ringlet, the difference is noticeably larger at the edges, where,
therefore,  
substantial collisional effects are needed there to produce a balance.
 This occurs because self gravity is significantly
amplified at a sharp edge where it overcompensates for the effect of differential 
precession, requiring a significant enhancement of collisional effects
to provide a balance. 

The reason for the amplified effects of self gravity
at an edge is that the tendency  for gravitational effects due to interior material to
cancel those arising from exterior material at any point in question is  absent.
This leads to an enhancement factor $\sim (1/2)\ln( W/W_{edge}),$ where
$W_{edge}$ is the characteristic scale associated with the edge, typically this factor $\sim 2-3,$
for the models considered here.

On account of this it is of interest to highlight solutions  for which the differential
precession is {\it exactly} compensated  for by  self-gravity in the central region of the ringlet,
but  then even larger collisional contributions are implied in the regions
close to the edges, an effect that is 
illustrated in
Fig.~\ref{figCS11b}, for both of the assumed values of $J_2$.

\begin{figure}
\begin{center}
\includegraphics[width=6cm,angle=270]{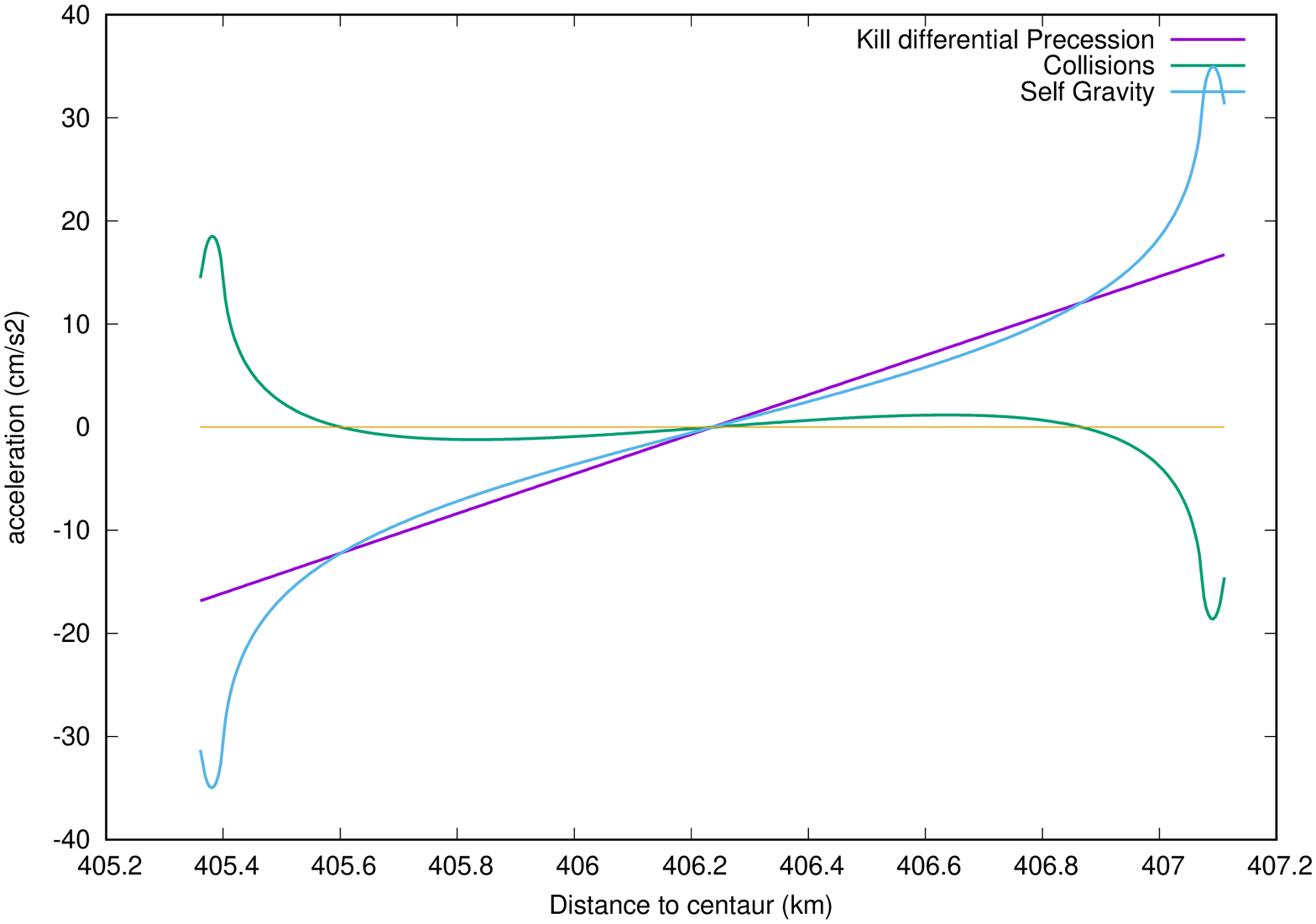}
\includegraphics[width=6cm,angle=270]{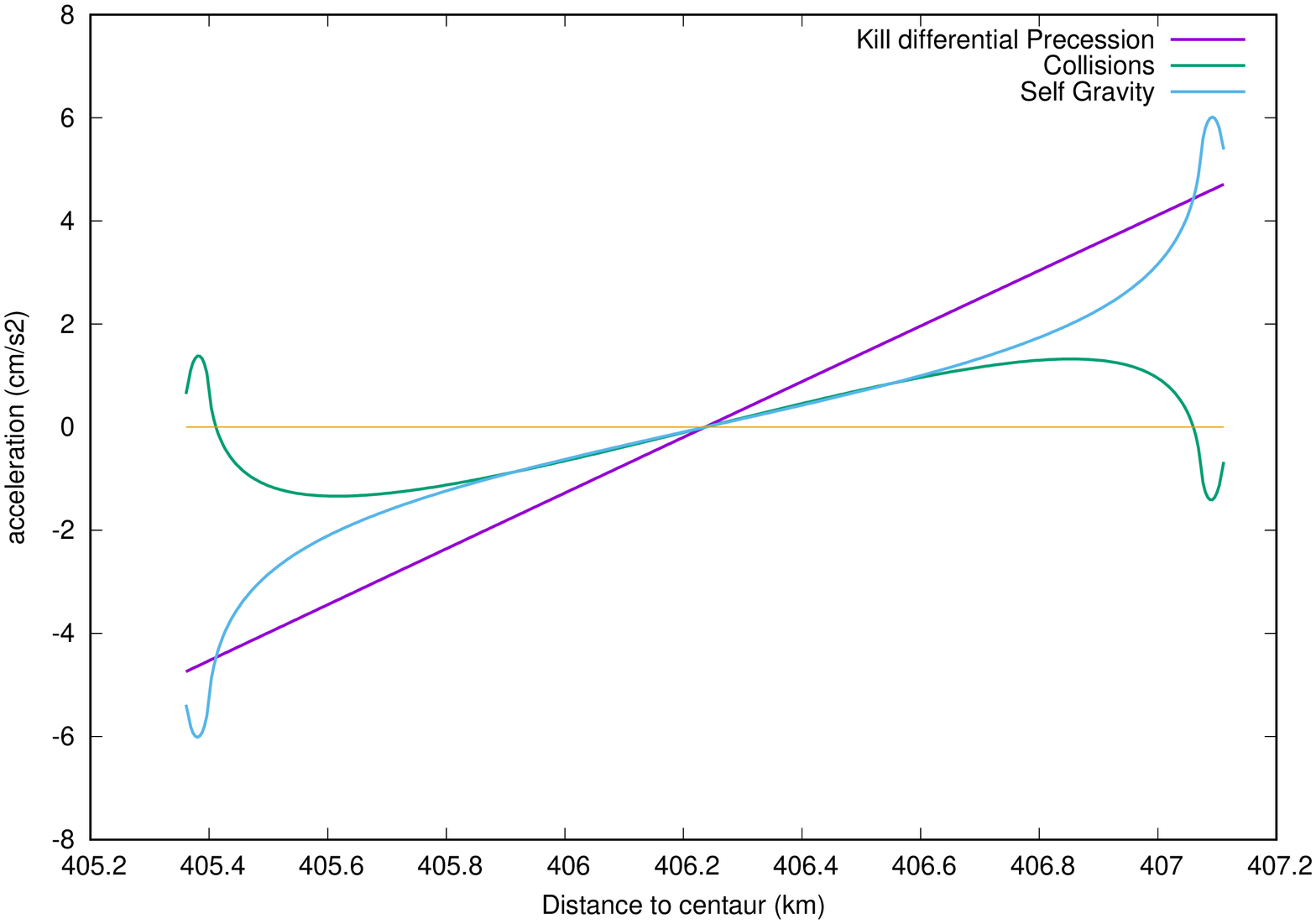}
\caption{Above: 
 The acceleration needed to cancel the differential precession
due to the oblateness of  (10199) Chariklo (magenta curve),
the acceleration due to self gravity (blue curve)
and that due to azimuthally integrated collisional effects (see equation (\ref{eqFc}))  (green curve)
for the CR2 ringlet model that minimises
the integrated collisional term, $F$,  for a value of 
$J_2^a=0.28$.
Below: As above but for a value of
 $J_2^b=0.076$.  For the values of  surface
  density and eccentricity gradient used to produce each plot
  see table \ref{tab:mins}.
 }
\label{figCS2}
\end{center}
\end{figure}

\subsubsection{The CR2 ringlet}\label{CR2}

We now apply the above analysis  
to the outer, narrower and
fainter ringlet CR2. We assume a mean distance from (10199) Chariklo
of $404 km$ and  the same value of the inner eccentricity  as for CR1. 
We  also assume  similar surface  density profiles  and  constant values for the
 parameter $q$, as
in the case of the  CR1 ringlet.  The central uniform   surface density
values explored also range between
$10 g cm^{-2}$ and $610 g cm^{-2}$ and as before, $q$ was taken to be
in the range $0.01 - 0.61.$ 
The resolution of the grid used for numerical modelling in this case
was approximately $2.2m.$ 


%

In the case of the CR2 ringlet the  integrated collision term is 
minimised for much smaller  values of the eccentricity gradient,
giving  $q=0.06$ for  $J_2^b=0.28$ and $q=0.01$ for  $J_2^b=0.076$,  
with a value of the surface density  in the first case equal to that for the corresponding
 CR1 case of $\Sigma = 10 g cm^{-2}.$   
As expected, the integrated
collisional contribution is significantly smaller as compared to   corresponding cases for  the
larger ringlet. The distributions of the accelerations for these cases are
illustrated in Figure \ref{figCS2}.

\begin{figure}
\begin{center}
\includegraphics[width=6cm,angle=270]{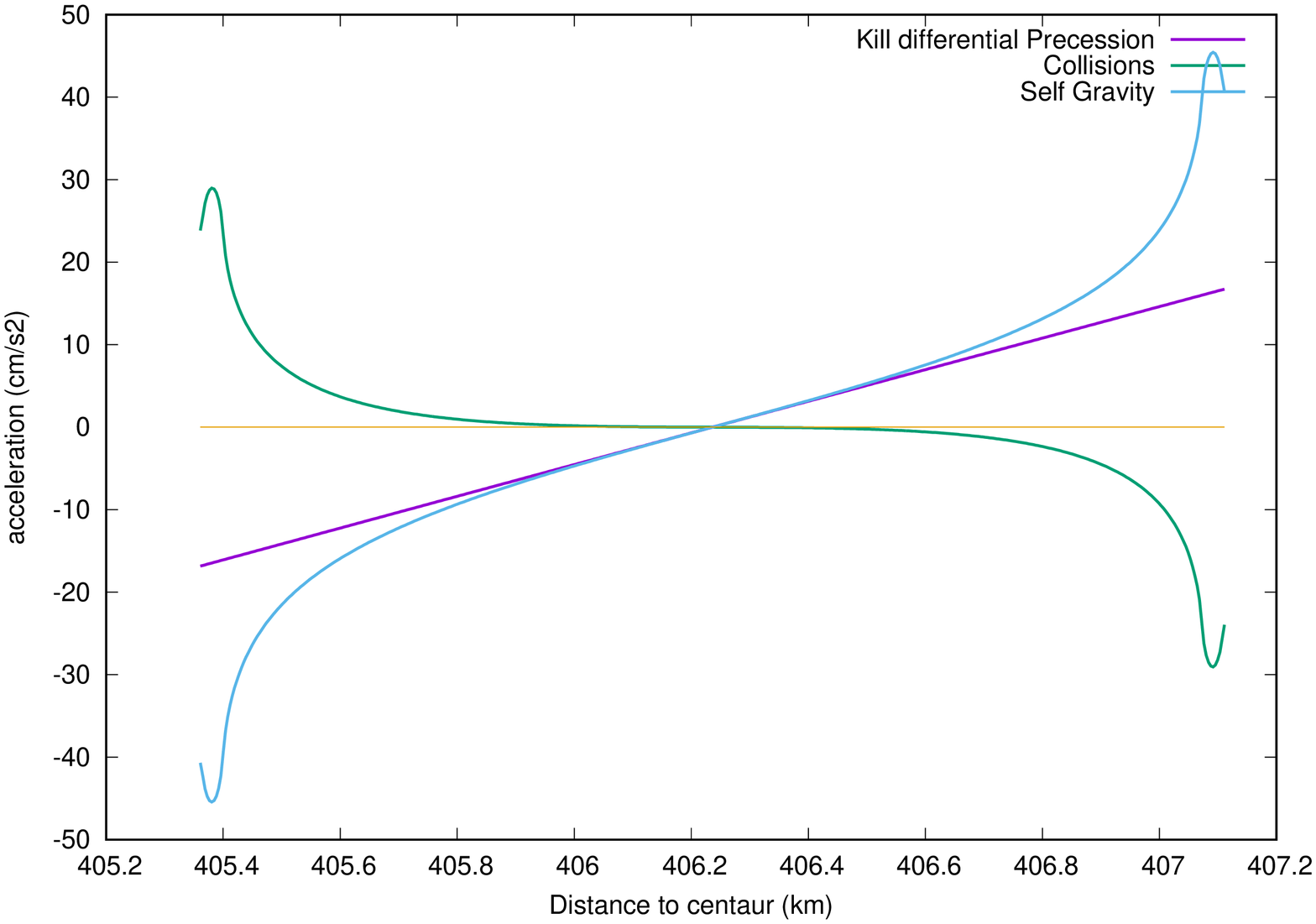}
\includegraphics[width=6cm,angle=270]{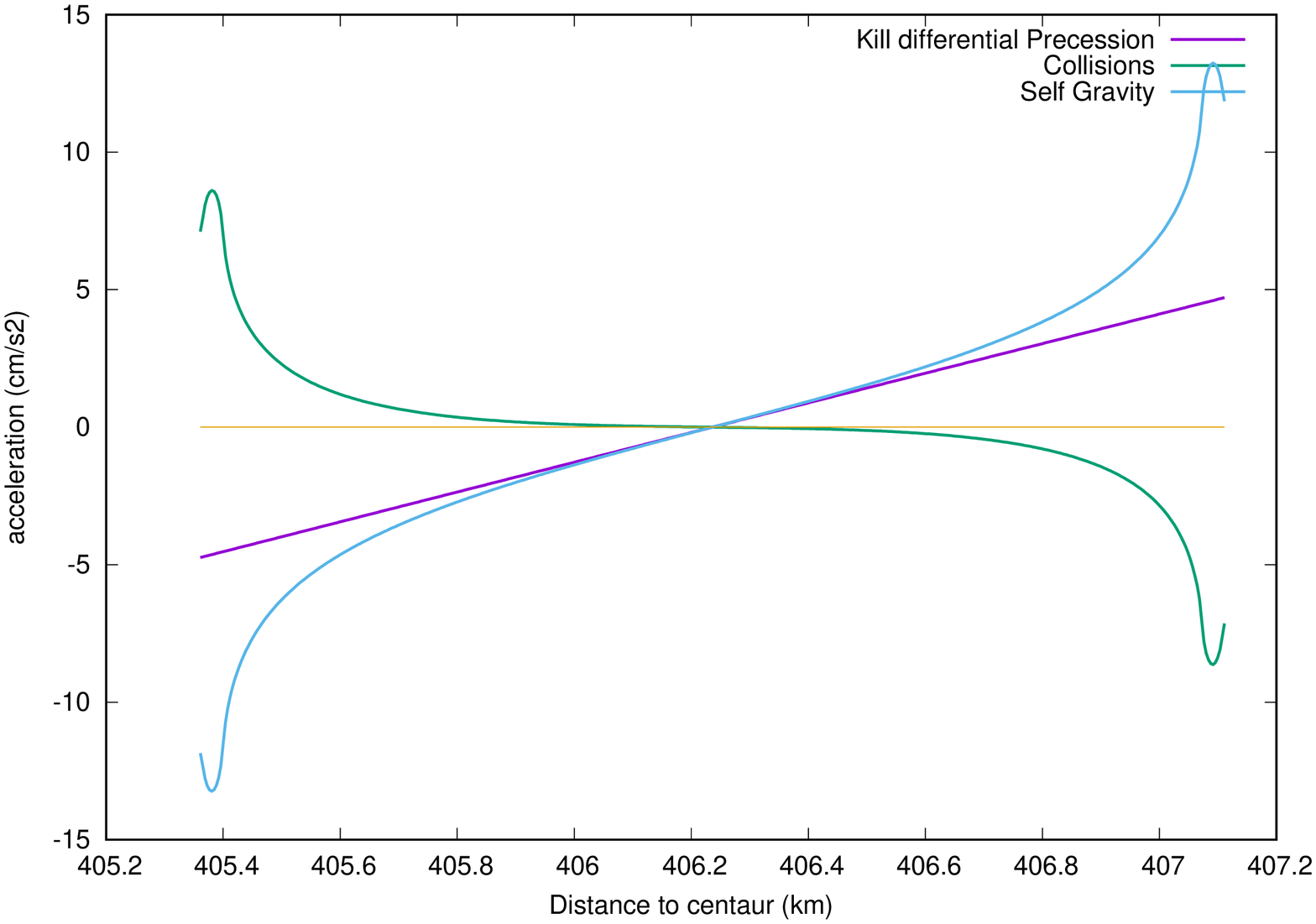}
\caption{
The acceleration needed to cancel the differential precession
due to the oblateness of (10199) Chariklo (magenta curve),
the acceleration due to self gravity (blue curve)
and that due to azimuthally integrated collisional effects (see
equation (\ref{eqFc}))  (green curve)
for the CR2 ringlet model, that practically removes the need for 
the collisional term in the middle section of the ring, for a value
of $J_2^a=0.28$ (above) and $J_2^b=0.076$ (below).  For the values of
surface
density and eccentricity gradient used to produce each plot
 see table \ref{tab:mins}.}
\label{figCS22b}
\end{center}
\end{figure}

As for the CR1 case we determined   models, that  almost entirely remove the need for 
the collisional term in the middle section of the ring. The parameters are given in 
table \ref{tab:mins}
and the distributions of the accelerations are
illustrated in Figure \ref{figCS22b}.

%
%
%

%
 
Qualitatively, the behaviour of the CR2 ringlet is similar that of the CR1 ringlet, i.e.
the acceleration imparted by self-gravity {\it approximately} balances the
differential precession, except at the edges, where the
self-gravity is dominant and significant forces arising from collisions are
needed to enable  rigid precession there.  

 Having made the above calculations which indicate the necessity 
of important collisional effects near the edges if they are sharp,
we remark that the  CR2 ringlet is of low optical depth (see Table~\ref{tab.2})
making collisional effects due to enhanced packing potentially implausible.
In this context we note that the edge profile is not constrained by the observations (B17)
so that it could be more tapered than assumed,  obviating the requirement for
enhanced collisional effects. This may be a consequence of there  being  no external
satellites driving ring eccentricity in this case.
\section{Discussion and Conclusions}

We have applied the theory of apse alignment in narrow
eccentric ringlets as considered in (PM05) and discussed further in
Section~\ref{sec.Theory} to the ring system of (10199) Chariklo.

We considered the conditions for the maintenance of rigid body
precession through the action of self-gravity and collisional effects
counteracting the effect of differential apsidal precession in
Sections~\ref{Rigidprec}-\ref{qval} going on to consider the
conservation of the ring radial action and the balance of collisional
dissipation and ring satellite torques required to maintain the
eccentricity in Sections~\ref{RA}-\ref{RST}.

A simple estimate for the collisional dissipation arising as a result
of close packing possibly occurring in an eccentric ring at a pinch
near pericentre \citep[see][]{DermottandMurray1980} or through
satellite ring interaction \citep{ChiangGoldreich2000} was given in
Section~\ref{CH_ring}.  Focussing on parameters appropriate to the 
optically thicker and more extensive CR1 ringlet, and using this
estimate, we showed in Section~\ref{viscd} that the mass of the 
satellites needed to prevent eccentricity decay associated with the
$m=1$ mode, for which the ringlet precesses uniformly, are estimated
to be similar to those needed to produce confinement against viscous
spreading as estimated in BR14.

In order to validate our procedures we applied the rigid precession
condition (Eq.\ref{pr5}) to the occultation data for the
$\epsilon$-ring of Uranus in Section~\ref{probU} and showed that a
balance occurs if the surface density corresponds to a canonically
estimated value of $30 g cm^{-2}$, which confirms the validity of our
theoretical considerations. Moreover, the result that the
accelerations resulting from enhanced inter-particle collisions at the
edges of the $\epsilon$-ring of Uranus act to increase the precession
rate at the inner edge and decrease it at the outer edge can be
understood with the help of a simple scenario, In this the eccentric
Keplerian orbits of particles in the neighbourhood of each of the
edges are modified by collisions occurring at an orbital phase near
pericentre with a hard wall located at the corresponding edge,
mimicking the effect of close packing.  It is easy to verify
that these act to increase the precession rate at the inner edge and
to decrease it at the outer one, in the same sense as the one induced
by the oblateness of the central object. 

Indeed, this result is qualitatively reproduced in the case of the
ringlets of (10199) Chariklo. Note that in the case of the optically
thin rings such as CR2 for which observations poorly constrain the
sharpness of the edges, close packing may only occur in localised
regions  near the edges close to pericentre if an eccentricity
is indeed excited by an external satellite.

We applied the rigid precession condition to the CR1 and CR2 ringlets
in Sections~\ref{CR1} and \ref{CR2}.  The value of the eccentricity
gradient, that minimised the collisional contribution, adopted in our
models for both the CR1 and the CR2 ringlets, ranges between $q
\approx 0.01$ and $q \approx 0.29$. The former is significantly
smaller  than the global estimate obtained assuming a balance
between the differential precession across the entire ring and effects
due to self-gravity as provided by equation~(\ref{valueq}).  That
gives a value of $q \approx 0.07$ , if $e=0.003$ and assuming that the
same set of parameters apply to both ringlets. However, it is
important to emphasise that such global estimates do not take into
account the increased contribution from self-gravity near sharp edges
that necessitates additional contributions from collisional effects as
has been seen in all our models.

For $J_2^{b}=0.076$
the most plausible surface densities independently
estimated  for the CR1 and CR2 ringlets of (10199)
Chariklo  have a ratio $\Sigma_{CR1} / \Sigma_{CR2}$, ranges
between $1$
and $2$, values that are
smaller  than the {\it observed} ratio of optical depths
$\tau_{CR1} / \tau_{CR2} \approx 9 $. On the other hand,  if a value of
$J_2^{a}=0.28$ is assumed, the estimated values of $\Sigma_{CR1}$ and  
$\Sigma_{CR2}$ are similar for very different values of $q$, which
give us a hint that, 
probably, the
real value of $J_2$ of (10199) Chariklo is smaller than this value. 
We remark that the ratio of $\tau_{CR1} / \tau_{CR2}$ observed is also approximately the ratio of the
widths of the CR1 and CR2 ringlets and so the ratio of surface density
to width is constant. This in turn implies that self-gravity is able
to approximately balance the differential precession across the
ringlets for the same values of inner eccentricity and  $q.$ 

 For both cases, the estimated values of $\Sigma$, ranging from 
$10 g cm^{-2} $ to $22 g cm^{-2} $ can correspond to plausible values of
the mean size of the ring particles. %
%
Since the
surface density and the optical depth are related  through
 $$ \Sigma =\frac{4}{3} \rho a \tau,$$
 if a bulk density of the ring particles is
about $1 g cm^{-3}$, the mean radius of the particles would range from $a
\approx 15 cm$ to $a \approx 33 cm$ for CR1.
For CR2,  $\Sigma = 10 g cm^{-2} $
would lead to  $a
\approx 150 cm.$ 

 Finally we remark that although the detailed discussion above
refers to the CR1 and CR2 ringlets of (10199) Chariklo for which we
assumed confinement and eccentricity maintenance was due to shepherd
satellites, the analysis of the maintenance of apsidal alignment
through the action of self-gravity and particle collisions described
in Sections \ref{sec.Theory} -\ref{qval} does not depend on this and
accordingly it should be applicable to planetary ringlets in general.
The energy and angular momentum input required to maintain the
eccentricity may arise from other sources
\citep[][]{Hedmanetal2010,Frenchetal.2016}.


\appendix

\section{Numerical solution of the self-gravity integral}
\label{ap1}

The integral of the self gravity term $g_d$ (Eq.~(\ref{gs2})) has a
singularity, when $r_0' \rightarrow r_0$,
which has to be dealt with by taking the principal value.
Therefore it cannot be
 calculated numerically in a straightforward manner because of the danger of
large errors being produced through evaluating the contribution to the integral
arising  close to
and at $r_0' = r_0.$ A way to
avoid this problem  is to  make a transformation
that enables evaluation of the integral via a 
discrete Fourier transform of the integrand  as follows.      
We  shall  assume that, as in the case of interest, 
the integral, ${\cal I}$,  an be transformed to  the general form
\be
{\cal I} = \int_{-1}^{1} \frac{G(x,x')}{x - x'}\ dx',
\label {intg}
\ee
with $(-1 \le x \le 1 )$ and $G(x,x')$ non singular.
We then perform a change of variable such that 
\be
x'= \cos \theta  \hspace{1cm} (0 \le \theta \le \pi ),
\label {intgb}
\ee
if we define $H(x,\theta)=G(x,\cos\theta)$, ${\cal I}$ is now written as:
\be
{\cal I}  = \int_{0}^{\pi} \frac{H(x,\theta)\sin \theta\ d \theta}{x - cos \theta}.  
\label {intg2}
\ee
Now, noting  that it vanishes for $\theta =0,$ and $\theta = \pi,$
 we express $H(x,\theta)$ as a Fourier sine  series of the form:
\be
H(x,\theta) = \sum_{n=1}^{+\infty} b_n\sin (n\theta)\label{Fourier},
\ee
where
\be
b_n(x) = \frac{2}{\pi} \int_{0}^{\pi} H(x,\theta)\sin(n\theta) d\theta.
\label{bn}
\ee
 Then from 
 Eq.~(\ref{intg2}) and Eq.~(\ref{Fourier}) we
can write
\be
{\cal I }= \sum_{n=1}^{\infty} b_n\ I_n,
\label{sumI}
\ee
where
\be
I_n =  \int_{0}^{\pi} \frac{\sin(n \theta)\sin \theta d\theta}{x -\cos \theta}.
\label {intgIn1}
\ee
 We note that the above integral  may be writen in the form 
\be
I_n =   \frac{1}{4}\ \int_{0}^{2\pi} \frac{(\cos((n-1) \theta)-\cos((n+1) \theta)) d\theta}{x -\cos \theta}.
\label{intgIn1a}\ee
To evaluate the   integral in (\ref{intgIn1a})
we consider it in the
complex plane  and apply Cauchy's  Theorem. To do this  we  introduce
the complex
variable $z=e^{{\rm i} \theta}$, where ${\rm i}$ is the usual imaginary unit.
The integral in  equation~(\ref{intgIn1a}) can be written as
\be
I_n = Re \left[ \frac{{\rm i}}{2} \oint_{{\cal C}} \frac{(z^{n-1}-z^{n+1})dz}{ (z-e^{i\ \alpha }) (z-e^{-i\ \alpha} ) }
 \right],
\label {intgIn1b}
\ee
where $x = \cos\alpha$ and ${\cal C}$ denotes the unit circle with the
singularities at $\exp(\pm {\rm i}\alpha)$ which lie upon it being
dealt with by taking the principal value.  By considering the contour
consisting of the unit circle with infinitesimally small semicircular
indentations at $\exp(\pm {\rm i}\alpha)$ and applying Cauchy's
theorem we can write

\be
I_n = Re \left[ \frac{{\rm i}}{2} \oint_{{\cal C'}} \frac{(z^{n-1}-z^{n+1})dz}{ (z-e^{i\ \alpha }) (z-e^{-i\ \alpha} ) }
 \right] +\pi\cos\alpha,
\label {intgIn1c}
\ee
where   ${\cal C'}$  is any closed curve interior to the  unit circle.
As there are no singularities inside it, the integral round it is zero and we have
\be
I_n = \pi \cos (n\alpha), \hspace{1cm} (\alpha \in [0,\pi]).
\label {intgIn20}
\ee
In any practical calculation the sum in  equation~(\ref{sumI}), being a Fourier series,  must be terminated  after a  finite
number of terms. To determine the optimal number  for
evaluation of  the self-gravity integral, we use the Sampling  Theorem.\\
We suppose that $H(x,x') \equiv H(x,\theta)$ is evaluated on an equally spaced grid on, $x' = [-1,1], $ with $n_{max}$ points\\
$x'_j = -1 +  2(j-1)/(n_{max}-1) , \hspace{2mm}  j= 1,2...n_{max}.$ 
The uniform grid spacing is $x'_{j+1}-x'_j = 2/(n_{max}-1).$ 
Corresponding to this are the unequally spaced grid points 
$\theta_j, \hspace{2mm}  j= 1,2...n_{max},$ which are such that $x'_j = \cos(\theta_j).$
If  $(\Delta\theta)_{min}$
corresponds to the period of the smallest scale that needs to be everywhere  resolved, 
we then require that
 $$ |\theta_j -\theta_{j-1}|  \le \frac{1}{2} (\Delta\theta)_{min},     j= 2, 3, ...n_{max}.  $$ 
\noindent The sampling frequency is then 
above the Nyquist frequency thus avoiding spurious effects that
arise from neglecting high frequency components. 

In practice we re-sample
the data, interpolating linearly between neighbouring points  as needed
to ensure that  the total number of points, $n_{max}$, is such that the condition 
$$   |\theta_j -\theta_{j-1}|  < \frac{1}{2} (\Delta\theta)_{min}$$ 
is satisfied for $j = 2,3...n_{max}.$ 
The final expression for the integral, $I, $  is then
\be
I = \pi\ \sum_{n=1}^{n_{max}} b_n\ \cos (n \alpha).
\label {intgIn2}
\ee
We recall that the coefficients $b_n$ must be obtained from Eq.~(\ref{bn}).
Considering that for a given value of, $x,$ we know the values of $H(x, \theta)$ at a discrete
number of  grid points, $\theta_j$, $j=1,2,3 ..n_{max}$, we approximate its
value at intermediate points  in $[\theta_{j-1},\theta_j]$ through
\be H(x,\theta) = A_j(x)\theta + B_j (x)\hspace{1cm} (\theta_{j-1} \le
\theta \le \theta_j), \label{bns}
\ee
where
$$ A_j = \frac{H(x,\theta_j) - H(x,\theta_{j-1})}{\theta_{j} - \theta_{j-1}},
$$
and 
$$ B_j = \frac{H(x,\theta_{j-1})\ \theta_j - H(x, \theta_{j})\ \theta_{j-1}  }{\theta_{j} - \theta_{j-1}}.
$$
We now perform the integral in (\ref{bn}) by summing contributions from each interval 
that can, after making use of (\ref{bns}),  be evaluated analytically and thus obtain
\begin{eqnarray}
&b_n = 2/\pi\sum_{j=2}^{n_{max}} 
A_j 
 \left(  \sin(n \theta_j) - \sin(n \theta_{j-1}) \right)/n^2   \nonumber\\
&+  2/\pi\sum_{j=2}^{n_{max}} A_j\left( \theta_{j-1}\ \cos(n \theta_{j-1}) - \theta_{j}\ \cos(n \theta_{j})  \right)/n
 \nonumber\\
&+ 2/\pi\sum_{j=2}^{n_{max}}B_j\left( \cos(n\theta_{j-1}) - \cos(n\theta_{j}) \right)/n.
\end{eqnarray}
We remark that although formally $b_n$ depends on, $x,$ for 
application to equation~(\ref{gs2}) for the special case when $q$ is
constant, as considered in this paper, this dependence appears through
a constant factor of the integrand in equation (\ref{gs2}) and so
$H(x,\theta)$ may effectively be regarded as independent of $x$ for
this special case of interest. In addition we remark for the examples
considered in this paper, the  radial edges  of the rings are not
resolved observationally. In the theoretical models the edges are not
resolved to within the grid spacing in $x'.$ As a consequence of this
integrals such as that  in  (\ref{bn}) are not accurate for, $x,$
being within a few grid points of the edges and should not be
evaluated there.

\newpage

\vspace{5cm}

\centerline{{\Large{\bf References}}}

\bibliographystyle{elsarticle-harv} 
\bibliography{Rings_Chariklo.fin}







\end{document}